\documentclass[11pt, 
abstract=true, 
captions=tableheading, 
]{article}

\usepackage[right=3cm, bottom=3.5cm]{geometry}

\usepackage[utf8]{inputenc} 
\usepackage[T1]{fontenc} 
\usepackage[main=english, ngerman]{babel} 

\usepackage{siunitx} 
\usepackage{graphicx} 

\usepackage{booktabs} 
\usepackage{multirow} 
\usepackage{csquotes} 
\usepackage{scrlayer-scrpage} 
\clearpairofpagestyles
\ihead{\headmark}
\ohead{\pagemark}

\usepackage[backend=biber, 
style=phys, 
articletitle=true, 
pageranges=false, 
language=english, 
biblabel=brackets 
]{biblatex} 
\usepackage{amsmath} 
\ifluatex
\usepackage{unicode-math} 
\fi

\usepackage{amsfonts, amsthm}

\theoremstyle{definition}

\theoremstyle{remark}

\usepackage{physics} 
\usepackage{aligned-overset}
\usepackage{cancel}

\usepackage{caption} 
\usepackage{subcaption} 
\DeclareCaptionSubType*{figure}

\usepackage{overpic}

\usepackage{hyperref} 
\usepackage{placeins} 

\usepackage{todonotes}
\usepackage{dsfont}
\usepackage{amssymb}
\usepackage[normalem]{ulem}

\usepackage{fontawesome}
\usetikzlibrary{backgrounds}
\usetikzlibrary{graphs}
\usetikzlibrary{positioning}
\usetikzlibrary{arrows.meta}
\usetikzlibrary{calc}
\usepackage{pgfplots}
\pgfplotsset{compat=1.9}
\usepackage{pgfplotstable}

\usepackage{algorithm}
\usepackage[pict2e, english]{struktex} 

\newcommand{\R}{\mathbb{R}}

\newcommand{\N}{\mathbb{N}}

\newcommand{\Prob}{\mathds{P}}

\definecolor{mydarkgreen}{HTML}{0b5313}
\definecolor{mycoolred}{HTML}{a4044d}
\definecolor{mylightgreen}{HTML}{50d075}
\definecolor{mydarkblue}{HTML}{1f4196}
\definecolor{myturquoise}{HTML}{26cdca}
\definecolor{mygreenishblue}{HTML}{256676}

\definecolor{c_navy}{RGB}{0, 0, 128}
\definecolor{c_maroon}{RGB}{128, 0, 0}
\definecolor{c_orange}{RGB}{245, 130, 48}
\definecolor{c_yellow}{RGB}{255, 225, 25}
\definecolor{c_lavender}{RGB}{220, 190, 255}
\definecolor{c_blue}{RGB}{0, 130, 200}
\definecolor{c_teal}{RGB}{0, 128, 128}

\usepackage{marvosym}
\usepackage{lscape}

\usepackage{authblk}

\newcommand\nnfootnote[1]{%
  \begin{NoHyper}
  \renewcommand\thefootnote{}\footnote{#1}%
  \addtocounter{footnote}{-1}%
  \end{NoHyper}
}

\makeatletter
\newcommand\colorwave[1][blue]{\bgroup \markoverwith{\lower3.5\p@\hbox{\sixly \textcolor{#1}{\char58}}}\ULon}
\font\sixly=lasy6 
\newcommand\smallO{
  \mathchoice
    {{\scriptstyle\mathcal{O}}}
    {{\scriptstyle\mathcal{O}}}
    {{\scriptscriptstyle\mathcal{O}}}
    {\scalebox{.7}{$\scriptscriptstyle\mathcal{O}$}}
  }

\makeatother

\begin{document}

\title{Extreme Value Analysis for Finite, Multivariate and Correlated Systems with Finance as an Example}
\author[]{Benjamin Köhler}
\author[]{Anton J. Heckens}
\author[]{Thomas Guhr}
\affil[]{\textit{\small Fakultät für Physik, Universität Duisburg--Essen, Duisburg, Germany}}
\date{\today}
\maketitle

\nnfootnote{\textit{Email addresses:} benjamin.koehler@uni-due.de (Benjamin Köhler), anton.heckens@uni-due.de (Anton J. Heckens), thomas.guhr@uni-due.de (Thomas Guhr)}
\vspace{-1cm}
\begin{abstract}
Extreme values and the tail behavior of probability distributions are essential for quantifying and mitigating risk in complex systems of all kinds. In multivariate settings, accounting
for correlations is crucial. Although extreme value analysis for infinite correlated systems remains an open challenge, we propose a practical framework for handling a large but finite number of correlated time series. We develop our approach for finance as a concrete example but emphasize its generality.
We study the extremal behavior of high-frequency stock returns after
rotating them into the eigenbasis of the correlation matrix. This separates and extracts various collective effects, including information on the correlated market as a whole and on correlated sectoral behavior from idiosyncratic features, while allowing us to use univariate tools of extreme value analysis. This holds even for high-frequency data where discretization effects normally complicate analysis. We employ a peaks-over-threshold approach and thereby fully avoid the analysis of block maxima.
We estimate the tail shape of the rotated returns while explicitly accounting for nonstationarity, a
key feature in finance and many other complex systems. Our framework facilitates tail risk estimation relative to larger trends and intraday seasonalities at both market and sectoral levels.
\end{abstract}

\section{Introduction}
Estimating properties of rare and extreme events, such as their magnitude, frequency and temporal dynamics is crucial for understanding systemic stability as well as for managing and mitigating risk in complex systems, ranging from weather, climate and hydrology to socio-economic systems such as traffic, social insurance and financial markets \cite{GeneralPerspectiveWeather2011, ClimateChange2014, HydrologyPaper2004, SafetyEstimation2006, Koedijk1990, EmbrechtsModellingExtremalEvents1997}. Complex systems often exhibit significant correlations, both in the time domain as well as over different system constituents. In finance, the latter gave rise to the method of diversification and Markowitz' famous Portfolio optimization \cite{Markowitz1952}. Long range dependency in absolute values of financial returns is well-established, see for instance Refs.~\cite{Ding1993, Lux1996}. In addition, nonstationarity is typically a key feature, since system dynamics often change drastically and extreme events such as market crashes occur \cite{Münnix2011,Münnix2012}. Recent empirical and analytical studies have shown that in general, fluctuations of correlations between stock returns lift the tails of the multivariate return distributions \cite{GuhrSchell2021, Manolakis_I, Heckens2025_II}. Thus analysis without an adequate handling of dependencies and using tools which rely on stationarity might lead to false risk assessment, in the worst cases to considerable underestimation of risk \cite{Salmon2012, Livan2012}.\par
The key theorems of extreme value theory for univariate time series were first derived by Fisher and Tippett \cite{FisherTippett1928} and formalized by Gnedenko \cite{Gnedenko1943}. Decades later, the mathematically equivalent description of extremal limit distributions arising from excess distributions was pioneered by Balkema, de-Haan \cite{BalkemaDehaan1974} and Pickands \cite{Pickands1975}. The field is application-driven and interdisciplinary in nature. Some areas where even early extreme value theory proved useful for empirical analysis, are hydrology, especially flood research, meteorology and material science \cite{GumbelReturnPeriod1941, Gumbel1942, Weibull1939}. Since then various fields have influenced mathematical research on a more general understanding of extreme values. All these results apply to the univariate case.\par
Extreme value theory for correlated systems has been studied in detail \cite{Majumdar2024, Majumdar2020Review}, with a focus on autocorrelations in a single time series, \textit{i.e.} in a univariate setting. Here we address an issue of conceptual and practical interest, namely extreme value statistics for a large, but finite number of cross-dependent time series, whose mutual bilinear statistical dependencies are captured by a Pearson correlation matrix. However, the measure of mutual dependencies is not limited to linear correlation measures, nonlinear measures such as tail-dependence matrices are also relevant. The goal of this paper is to address the following four points:
\begin{itemize} 
\item Develop a framework for analyzing extreme values in multivariate, highly correlated time series, using tools from univariate extreme value statistics while still capturing properties of the full correlated system.
\item Introduce and motivate the rotation into the eigenbasis of the correlation matrix as a practical way to handle dependence and to identify collectivity.
\item Apply a threshold-based approach, which naturally yields tail behavior as well as extreme value statistics. This allows us to avoid common, but rather arbitrary block maxima approaches.
\item Discuss how to account for nonstationarity and what this implies for our method.
\end{itemize}
To be concrete, we use finance as an example and carry out an extreme value analysis of empirical data from the New York Stock Exchange (NYSE) for the year 2014 at an intraday resolution level. We apply a method focusing on the tail behavior of the marginal distributions, \textit{i.e.} the tail behavior without accounting for seasonality and nonstationarity, and compare with the results of our new method incorporating intraday volatility clustering and nonstationarity. Our approach allows analyzing high-frequency returns on a timescale of seconds. \par
This contribution is meant to be primarily a conceptual study. We develop novel
approaches and methods and present a proof of concept. Detailed applications to
particular systems and a practical implementation, e.g. for portfolio optimization, as well as other comprehensive empirical analyses would require large-scale studies beyond the scope of this conceptual work. The concepts and approaches presented in this paper provide plenty of material for further research.
\par
In Sec.~\ref{s:extremevaluetheory}, the aim is to give a short and certainly non exhaustive introduction to extreme value analysis, regarding independent as well as weakly serial dependent random variables. In Sec.~\ref{s:datamethods} we introduce our estimation methods, as well as the data set and data transformation. We describe the rotation of the returns and investigate the interpretation of these rotated returns in Sec.~\ref{s:multivariate_setting}. Finally, Sec.~\ref{s:dataanalysisresults} and Sec.~\ref{s:local_estimation} portray the main results of our analysis, conclusions are given in Sec.~\ref{s:conclusion}.

\section{Univariate Extreme Value Theory \label{s:extremevaluetheory}}
This short depiction of extreme value theory (EVT) is mainly based on Refs.~\cite{haan2006, Coles2001}. We focus on maxima, but the theory is derived analogously for minima. In Sec~\ref{ss:iid}, we summarize the basic theory for stationary, independent and identically distributed series of random variables. In Sec.~\ref{ss:dependent} we deal with the generalization to processes with serial dependence, that is dependence of the random variables in time and in Sec.~\ref{ss:extremalindex} we briefly describe the extremal index.

Importantly, we afterwards study a finite, multivariate and correlated system, in our data analysis. There are methods of multivariate extreme value analysis for time series analysis, which can be found for instance in Ref.~\cite{haan2006}. However, we provide a new method of analyzing such a system, relying solely on basic univariate EVT and we prove its usefulness for our application.

\subsection{Independent Random Variables \label{ss:iid}}
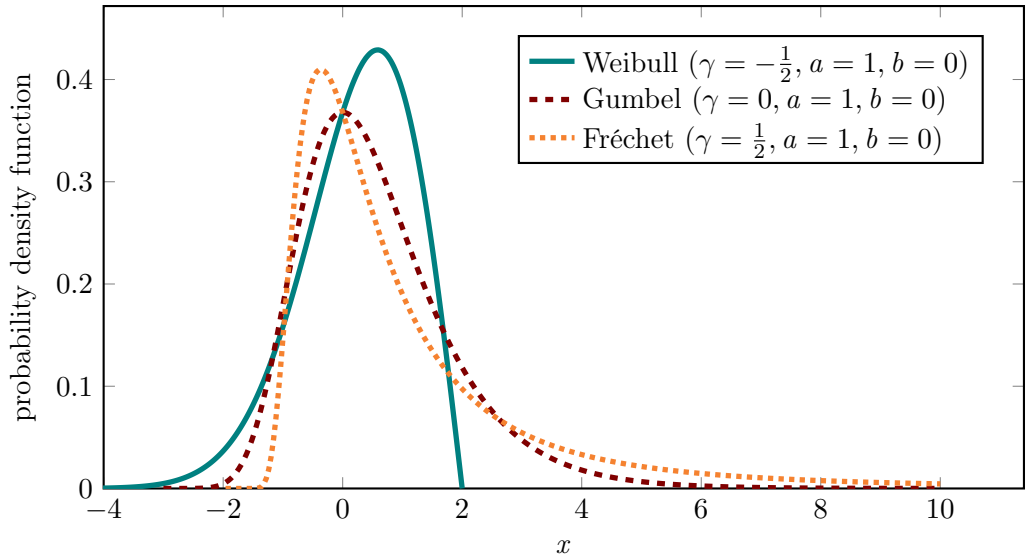
\begin{figure}
    \begin{tikzpicture}
    \begin{axis}[
        width=0.9\textwidth,
        height=8cm,
        xlabel={$x$},
        ylabel={probability density function},
        legend style={at={(0.45,0.94)}, anchor=north west},
        legend cell align={left},
        domain=0:10,
        xmin=-4,
        samples=300,
        thick,
        ymin=0,
    ]
    \addplot [c_teal, domain=-5:2, line width=2pt]
        { (1-0.5*x)^(-(1-1/0.5)) * exp( -(1-(0.5)*x)^(+1/0.5) ) };
    \addlegendentry{Weibull ($\gamma = -\frac{1}{2}, a=1, b=0$)}
    
    \addplot [c_maroon, dashed, domain=-3:10, line width=2pt]
        { exp(-x)*exp( -exp(-x))};
    \addlegendentry{Gumbel ($\gamma = 0, a=1, b=0$)}
    
    \addplot [c_orange, dotted, domain=-2:10, line width=2pt]
        { (1+0.5*x)^(-(1+1/0.5)) * exp( -(1+(0.5)*x)^(-1/0.5) ) };
    \addlegendentry{Fréchet ($\gamma = \frac{1}{2}, a=1, b=0$)}
    
    \end{axis}
    \end{tikzpicture}
    \caption{Probability density function $g^{\text{EV}}_{\gamma, b, a}(x)$ for $\gamma=0$ and special cases $\gamma=-1/2<0, \gamma=1/2>0$.}
    \label{fig:gev_density}
\end{figure}

The Fisher-Tippett-Gnedenko theorem \cite{FisherTippett1928, Gnedenko1943} addresses the maxima of a univariate series $(X_i)_{i\in\N}$ of independent and identically distributed (i.i.d.) random variables. It states that if there are series of rescaling constants $a_n \in \R_{>0}, b_n \in \R$, so that the limit of the cumulative distribution function of $(\max_{i=1,\ldots,n} X_i - b_n)/a_n$ exists then it is one of three types of distributions. Mathematically, the cumulative distribution function can be found as
\begin{align}
    \Prob\qty( \frac{\max_{i=1, \ldots, n} X_i - b_n}{a_n} \leq x) &= \Prob\qty(\bigcap_{i=1, \ldots, n} \qty{X_i \leq a_n x + b_n})  \notag\\
    &=\prod_{i=1}^n \Prob(X_i \leq a_n x + b_n) = F^n(a_n x + b_n)\, , \label{eq:cdfextremes_calc}
\end{align}
where the second and the third equality signs follow from independence and identity of distribution, respectively. The theorem now states that if $\lim_{n\to\infty} F^n(a_n x + b_n)$ exists in a nondegenerate way, then
\begin{align}
    \lim_{n\to\infty} F^n(x) = G^{\text{EV}}_{\gamma, b, a}(x) = \begin{cases} \exp(-\qty(1+\gamma \qty(\frac{x-b}{a}))^{-1/\gamma})\,, & \gamma \in \R\setminus \qty{0} \\
    \exp(-\exp(-\frac{x-b}{a}))\,, & \gamma = 0
    \end{cases} \label{eq:GEV}
\end{align}
with constants $a\in \R_{>0}, b\in\R$. The most relevant parameter is $\gamma$, since it distinguishes the three fundamental types of behavior of extreme values. The cumulative distribution is of Gumbel, Fréchet or Weibull type for $\gamma=0$, $\gamma>0$ and $\gamma<0$, respectively. The corresponding probability density function reads
\begin{align}
    g^{\text{EV}}_{\gamma, b, a}(x) = \frac{1}{a} \begin{cases} \qty(1+\gamma(\frac{x-b}{a}))^{-1 + 1/\gamma} \exp(-\qty(1+\gamma \qty(\frac{x-b}{a}))^{-1/\gamma})\,, & \gamma \in \R\setminus \qty{0} \\
    \exp(-\frac{x-b}{a})\exp(-\exp(-\frac{x-b}{a}))\,, & \gamma = 0
    \end{cases}\label{eq:GEV_density}
\end{align}
and is depicted in Fig.~\ref{fig:gev_density}.
While the Gumbel distribution has unbounded support on the real axis and decays exponentially in the tails, the Fréchet distribution has a fixed lower limit and describes a power law decay in the upper tail. On the other hand, the Weibull distribution has an upper limit and a power law in its lower tail. Importantly the series $a_n, b_n$ are not necessarily unique. If an underlying cumulative distribution $F$ is known there are however strategies to calculate specific series if they exist, see for instance Ref.~\cite{Majumdar2024}. 

The common method to analyze extreme value behavior in measured data is the block maxima (BM) approach. The time series is divided into blocks from which the maxima are extracted, and $G^{\text{EV}}_{\gamma,b,a}$ is then fitted to the distribution of maxima. The parameters $\gamma$ and the appropriate rescaling in the limit $a, b$ are typically estimated from the fit. For an exemplary application of the BM approach in a financial setting, see Ref.~\cite{Longin1996}.

\begin{figure}
    \centering
    \begin{tikzpicture}
    \begin{axis}[
        width=0.9\textwidth,
        height=8cm,
        xlabel={$x$},
        ylabel={probability density function},
        legend style={at={(0.55,0.94)}, anchor=north west},
        legend cell align={left},
        domain=0:5,
        xmax=6,
        xmin=-0.1,
        samples=300,
        thick,
        ymin=0,
    ]
    \addplot [c_teal, domain=0:10, line width=2pt]
        { (1-0.5*x)^(-(-0.5+1)/-0.5) };
    \addlegendentry{GPD ($\gamma = -\frac{1}{2}, \sigma=1$)}
    
    \addplot [c_maroon, dashed, domain=0:10, line width=2pt]
        { exp(-x)};
    \addlegendentry{GPD ($\gamma = 0, \sigma=1$)}
    
    \addplot [c_orange, dotted, domain=0:10, line width=2pt]
        { (1+0.5*x)^(-(+0.5+1)/+0.5) };
    \addlegendentry{GPD ($\gamma = \frac{1}{2}, \sigma=1$)}
    \end{axis}
    \end{tikzpicture}
    \caption{Probability density function $g^{\text{GPD}}_{\gamma, \sigma}(x)$ for $\gamma=0$ and special cases $\gamma=-1/2<0, \gamma=1/2>0$.}
    \label{fig:GPD_density}
\end{figure}
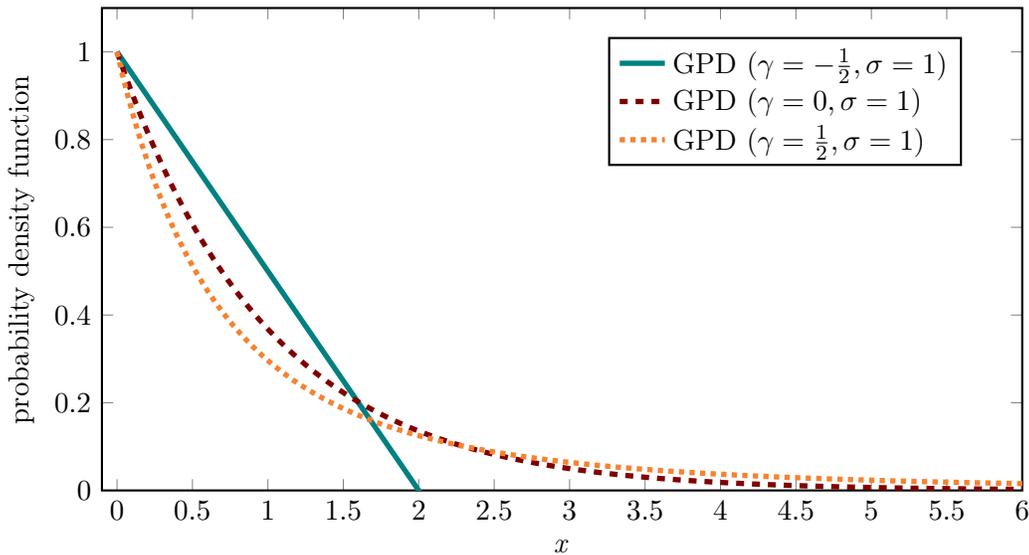

The Pickands--Balkema--DeHaan theorem \cite{BalkemaDehaan1974, Pickands1975} paves the road for a quite different method of analysis. The focus is on the whole tail of the marginal probability distribution for a series of i.i.d. random variables. We look at the cumulative conditional excess distribution function, defined as
\begin{align}
    F_{u}(x) = \Prob(X-u \leq x \, \vert\,  X>u) = \frac{F(x+u) - F(u)}{1-F(u)}\, .\label{eq:conditionalexcessdf}
\end{align}
It describes the cumulative distribution of random variables, conditioned on their exceedance of a threshold $u$.
The theorem states, that if there is an appropriate rescaling such that $\lim_{n \to\infty} F^n(a_n x + b_n)$ exists as in the premise of the Fisher--Tippett--Gnedenko theorem, then
\begin{align}
    \lim_{u\to\infty} F_u(x) = G^{\text{GPD}}_{\gamma,\sigma}(x)\,, \qqtext{where} G^{\text{GPD}}_{\gamma, \sigma}(x) = \begin{cases} 1 - (1+\frac{\gamma x}{\sigma})^{-1/\gamma}\, , & \gamma \in \R\setminus\qty{0} \\ 1 - e^{-\frac{x}{\sigma}}\, , & \gamma=0\end{cases}\, . \label{eq:GPD}
\end{align}
This family $G^{\text{GPD}}_{\gamma,\sigma}$ is called the generalized Pareto distribution (GPD). The parameter $\gamma$ is the same as in the family of extreme value distributions \eqref{eq:GEV}, while
\begin{align}
    \sigma = a + \gamma(u-b)\,\label{eq:linearparameterrelation}.
\end{align}
The corresponding probability density function reads
\begin{align}\label{eq:GPD_density}
    g^{\text{GPD}}_{\gamma,\sigma}(x) = \frac{1}{\sigma} \begin{cases} (1+\frac{\gamma x}{\sigma})^{-1/\gamma - 1}\, , & \gamma \in \R\setminus\qty{0} \\ e^{-\frac{x}{\sigma}}\, , & \gamma=0 \end{cases}
\end{align}
and is depicted in Fig.~\ref{fig:GPD_density}.

This result leads to another technique of empirically extracting extremal behavior, called the peaks-over-threshold (POT) approach. We choose a threshold $u$ and estimate the parameters $\gamma,\sigma$ from fitting $G^{\text{GPD}}_{\gamma,\sigma}$ to the empirical cumulative conditional excess distribution function $F_{u}^{\text{empirical}}(x)$.

There are several practical advantages of the POT approach over the BM estimation \cite{Coles2001}. While the latter takes into account only one data point in each block and discards the others, the former uses all exceedances over a certain threshold. Thus POT estimations often make more efficient use of the available data, which is crucial when dealing with extremes which are rare events and often leads to faster convergence in data analysis contexts. Due to the use of higher-order statistics instead of only first-order statistics, the POT approach is also arguably more intuitive in fields where not only the extremes but the whole tails of distributions are of general interest. Furthermore, although relying on a threshold, the POT are not tied to a certain block size in the time domain. Different block sizes can bias results differently, depending on the dynamics of the underlying process. For exemplary applications of POT we refer to Refs.~\cite{Coles2001, Hosking1987} and in particular regarding a financial setting to Ref.~\cite{McNeilFrey2000}.

\subsection{Dependent Processes \label{ss:dependent}}
For strongly correlated processes, the behavior of extremes remains poorly understood in general. Analytical results for exactly solvable systems can be found in Ref.~\cite{Majumdar2024}, while numerical investigations are discussed in Ref.~\cite{Micciche2025}. However, the findings obtained for the independent case can be generalized to dependent, but still stationary series of random variables under two types of restrictions usually denoted as $\textbf{D}$ and $\textbf{D}^\prime$ in the literature \cite{Leadbetter1974}.
\begin{itemize}
\item The first condition $\textbf{D}$ is a \textit{weak} mixing condition. For $u_n = a_n x + b_n$, there must be a counting sequence $l_n = \smallO(n)$, such that for the two sets of disjoint indices $A\subset\qty{1, \ldots, k}, B\subset\qty{k+l_n, \ldots, n}$ one has
\begin{align*}
    \max_{A,B}\qty{\vert \Prob(X_i \leq u_n, i\in A \cup B) - \Prob(X_i \leq u_n, i \in A)\Prob(X_i \leq u_n, i \in B) \vert} \xrightarrow[]{n\to\infty} 0\, .
\end{align*}
In the independent case the above difference is zero by definition. For dependent processes this means that for growing distances between two disjoint sets of subsequent random variables, the random variables are decreasingly dependent on each other. This is a notion of long-range independence. In the physics literature a similar condition is often used, by stating that there is a finite correlation length beyond which the variables become uncorrelated, see for instance Ref.~\cite{Majumdar2020Review}. 
\item The second, \textit{stronger} mixing condition $\textbf{D}^\prime$ prohibits short-range clustering,
\begin{align*}
    \lim_{n\to\infty} \qty(n \sum_{i=2}^{\lfloor n/k \rfloor} \Prob(X_1>u_n, X_i >u_n)) \xrightarrow[]{k\to\infty} 0\,,
\end{align*}
by restricting the joint probability of two extremes occurring in an ever shorter time window $\lfloor n/k \rfloor$. The sum accounts for all pairs in this time window.
\end{itemize}
If $\textbf{D}$ is satisfied, then in the limit of $n\to\infty$, the value of $\gamma$ is the same as for a series of independent variables with the same marginal distribution and only the appropriate rescaling differs, see Ref.~\cite{Leadbetter1974} and compare Ref.~\cite{Coles2001} for a more intuitive derivation in the Fisher--Tippett--Gnedenko setting. The key point is that one has to correctly estimate clustering of extremes in time. In empirical analysis, it is a standard approach to use declustering algorithms, where only the independent maxima of extremal clusters are used for analysis. However, in the mathematical limit both cumulative distributions, the one of the cluster maxima and the one of the clustered threshold exceedances converge to cumulative distributions with the same $\gamma$ when $\textbf{D}$ is satisfied, see also the contribution of Anderson \cite{DiscussionDavison1990} in the discussion of Ref.~\cite{Davison1990}. Yet, to make use of this equivalence one has to assume stationarity after all. In intraday financial data, nonstationarity and serial dependence is clearly a factor, which will be relevant later on in Sec.~\ref{s:dataanalysisresults} and Sec.~\ref{s:local_estimation}.\par
Another important caveat, when applying the above results to real data, is that in order to use the result under condition $\textbf{D}$, the degree of long-range serial dependence must be quantified carefully. A series may show no significant linear autocorrelations, but a nonlinear transformation may show autocorrelations and therefore the series is not independent. If these nonlinear autocorrelations are strongly persistent, above results may not be directly applicable in data analysis since $\textbf{D}$ is not satisfied. In this case, convergence to the theoretical distributions has to be thoroughly checked.

\subsection{Extremal Index\label{ss:extremalindex}}
If the stronger condition $\textbf{D}^\prime$ is satisfied, even the rescaling is the same as in the i.i.d case, see Ref.~\cite{Leadbetter1983}. However, for empirical time series of many systems, including finance, $\textbf{D}^\prime$ is not satisfied and we have to resort to $\textbf{D}$ because short range clustering is apparent. The strength of this clustering is usually captured by the extremal index $\Theta \in \qty[0,1]$, which is defined as the inverse of the mean cluster size of extremes. We briefly sketch an intuitive interpretation of the extremal index. In addition we relate the index to the waiting times between exceedances. For more details, mathematically exact definitions and proofs, we refer to the literature, for instance Refs.~\cite{OverviewExtremalIndex2019, FerroSegers2003, Leadbetter1983}. Given the total number of threshold exceedances $n$, the number of independent clusters of exceedances $n_C$ is approximately
\begin{align}\label{eq:number_clusters}
    n_C \approx n \Theta\, ,
\end{align}
where $\Theta$ is the inverse mean cluster size. The probability of $X_i$ exceeding $u$ is $1-F(u)$. An exceedance falls into a cluster with probability $1-\Theta$, but different clusters are independent of each other. Either the exceedance falls into a cluster, then a rescaled waiting time $w$ is zero, or it starts a new cluster and is independent from previous observations, then $w$ is larger than zero. Thus, the waiting times between exceedances $\tau_i$ rescaled with the probability of exceeding the threshold follow the probability density function \cite{FerroSegers2003}
\begin{align}\label{eq:prob_waiting_times}
    \lim_{u\to\infty}\Prob(\tau_i (1-F(u)) = w) = \begin{cases}
    \displaystyle 1-\Theta,& w = 0 \\
    \displaystyle \Theta\, \mathrm{Exp}(\Theta), & w>0
    \end{cases}\,,
\end{align}
where $\mathrm{Exp}(\Theta)$ denotes the probability density function of the exponential distribution with mean $1/\Theta$. In the case $\Theta = 1$, we retain the pure exponential density of waiting times, as expected for a Poisson process, there is no clustering. If we calculate the first two moments of the distribution of $\tau_i$ itself from \eqref{eq:prob_waiting_times}, we find
\begin{align}\label{eq:moments_waiting_times}
    \mathbb{E} [ \tau_i] = \frac{1}{1-F(u)}, && \mathbb{E}[ \tau_i^2 ] = \frac{2}{\Theta(1-F(u))^2},
\end{align}
and thus the identity \cite{FerroSegers2003}
\begin{align}\label{eq:ferro_segers_theory}
    \Theta = \frac{2(\mathbb{E}[\tau_i])^2}{\mathbb{E}[\tau_i^2]}
\end{align}
follows which only involves the first two moments of the waiting time distribution between exceedances.

\section{Data Processing and Methods of Analysis \label{s:datamethods}}
In Sec.~\ref{ss:data_processing} we describe our data and processing methods. We provide the statistical estimators and methods of estimation for the tail index $\gamma$ and the extremal index $\Theta$ in Sec.~\ref{ss:estimators}.

\subsection{Processing the Data \label{ss:data_processing}}
Our data consists of time series obtained for $K$ stocks from the Daily TAQ (Trade and Quote) files of the NYSE (New York Stock Exchange) in 2014. We choose an amount of $K=479$ liquid stocks from the S\&P500. The smallest provided time resolution is \SI{1}{\milli\second}, we choose a resolution of seconds for our analysis. Since market dynamics shortly after the opening and shortly before closing differ significantly from the main trading phase, we discard the first ten and last fifteen minutes of each trading day. Especially in the phase of the closing auction, dissemination of the closing auction order imbalance leads to frequent sharp and large spikes in volatility and volume \cite{BacidoreTradingClose2012}, which are only visible at high frequency. In the year 2014, the NYSE started this dissemination at 15:45:00 \cite{NYSERegulationRule}, therefore we exclude the trading phase after this time. For one full trading day we end up with $T_\text{day} = 21899$ seconds of data. The early closing days of 24/12/2014, 28/11/2014, 03/07/2014 
(dd/mm/Y) are removed completely, so that we end up with $N_\text{days}= 248$ full trading days in 2014. 
This leaves us with a total amount of $T=N_\text{days} T_\text{day} = 5430952$ seconds per stock.\par
Since trading is sometimes sparse at this high resolution, we construct the midpoint price for all stocks $k$ and all times $t$ by averaging best bid $b_k$ and best ask $a_k$ in the order book,
\begin{align}
    m_k(t) = \frac{a_k(t) + b_k(t)}{2}\,, \quad k=1,\ldots,K\,, \label{eq:midpoints}
\end{align}
and use the last available midpoint price in each second $t$. When no bids or asks are available in one second, midpoint prices are estimated as the last existing midpoint price in time. From this, we calculate the logarithmic returns
\begin{align}
    G_k(t, \Delta t) = \ln m_k(t+\Delta t) - \ln m_k(t) = \ln \frac{m_k(t+\Delta t)}{m_k(t)}\, , \label{eq:logreturns}
\end{align}
for all $k$ and $t$, where we denote the return horizon with $\Delta t$. We restrict ourselves to intraday trading, thus overnight jumps in price are not included as returns. After arranging the return time series as rows of the $K\times T$ data matrix $G$, we normalize the time series to zero mean and standard deviation of unity
\begin{gather}
        M_k(t) = \frac{G_k(t) - \langle G_k(t) \rangle_t}{\sigma_k}\, ,\label{eq:normalization}
\end{gather}
where $\langle G_k(t) \rangle_t$ and $\sigma_k^2 = \operatorname{var}_t(G_k(t))$ are sample mean and variance.

\subsection{Estimation Procedure \label{ss:estimators}}
\begin{figure}
    \begin{subfigure}{0.9\textwidth}
        \centering
        \begin{tikzpicture}
        \begin{axis}[
        axis lines=left,
        xlabel={$i$},
        ylabel={values $x_t$},
        xlabel style={at={(axis description cs:1.02,0)}, anchor=west},
        ylabel style={at={(axis description cs:0,1.0)}, anchor=south, rotate=-90},
        ymin=0, ymax=1,
        xmin=0, xmax=100,
        xtick={4,9,10,15,24,27,29,45,48,53,55,60,62,69,76,82,98},
        xticklabels={1,2,3,4,5,6,7,8,9,10,11,12,13,14,15,16,17},
        xticklabel style={
        yshift={-mod(\ticknum,2)*1em},
        },
        ytick={0.55},
        yticklabels={\color{c_maroon} threshold $u$},
        axis line style={thick},
        ytick style={draw=none},
        width=\textwidth,
        height=7cm,
        clip=false
        ]
        \foreach \x/\y in {1/0.5253198714088679, 2/0.23341971281509916, 3/0.017617344749315738, 4/0.6681567610447261, 5/0.024437017874249123, 6/0.4471605982116483, 7/0.199724013003651, 8/0.06625506773533595, 9/0.5887513739246971, 10/0.6450874534691479, 11/0.28059447065472687, 12/0.4916658748897756, 13/0.5083736737284791, 14/0.34993950033952503, 15/0.7506968652653274, 16/0.18781761326696847, 17/0.23529026261905905, 18/0.04637767363754076, 19/0.33757609097343083, 20/0.03039049735847787, 21/0.10267614778472312, 22/0.286530917552885, 23/0.008590545704922055, 24/0.6690877515131064, 25/0.044129761600371874, 26/0.17594416123117818, 27/0.9692493737440198, 28/0.2481168966594366, 29/0.9813158095201042, 30/0.38320009398223936, 31/0.34939292907330616, 32/0.07981076336229759, 33/0.47032145301342176, 34/0.3348003990530816, 35/0.5132599825254373, 36/0.31147347978292794, 37/0.19217722749143457, 38/0.0224709405135961, 39/0.31853973551858544, 40/0.38142987818881957, 41/0.45166883916048334, 42/0.48829235895416206, 43/0.11223368717111575, 44/0.5477149605604735, 45/0.6948329131045561, 46/0.5151267995962255, 47/0.43399584725175944, 48/0.729113455751621, 49/0.29200413672009157, 50/0.24139514252725902, 51/0.2176254699271167, 52/0.4536612047154135, 53/0.9125939032254404, 54/0.25286355654897846, 55/0.9903099777882203, 56/0.06931390748992629, 57/0.0006608249605391025, 58/0.3155178633779054, 59/0.33569455314146157, 60/0.9944987415039257, 61/0.15607907420876196, 62/0.5748709992295346, 63/0.5401715409156146, 64/0.4750296445147848, 65/0.3984607975602402, 66/0.49226491497782576, 67/0.5186879662926261, 68/0.0043435301166758365, 69/0.7746071623079815, 70/0.5389117310509783, 71/0.0454745540918483, 72/0.07775215091481588, 73/0.2699880145771954, 74/0.03303480230714859, 75/0.3954311604860138, 76/0.5973998012683518, 77/0.10370868774504435, 78/0.03545393710604023, 79/0.247446825103616, 80/0.2596288226017181, 81/0.16211095998560843, 82/0.928261484705258, 83/0.349478843754382, 84/0.5370841616889357, 85/0.24661546810190452, 86/0.08016591315093055, 87/0.5321830720090855, 88/0.11525952378724046, 89/0.1467982681790443, 90/0.22657556753006747, 91/0.2831777234416628, 92/0.13623503493324626, 93/0.16410078447327217, 94/0.004634811554663521, 95/0.30739731528253506, 96/0.2629491673958458, 97/0.22848666035997983, 98/0.7907698693079397, 99/0.36388588071943706, 100/0.408712868095583}{
            \edef\temp{
            \noexpand\draw[c_teal, thick] (axis cs:\x, 0) -- (axis cs:\x, \y);
            \noexpand\node[draw=c_teal, fill=c_teal, circle, inner sep=1pt] at (axis cs:\x,\y) {};
            }\temp
        }
        \addplot[dashed, c_maroon, thick, domain=0:12] coordinates {(0, 0.55) (100, 0.55)};
        \foreach \x/\t/\c in {4/5/1,  9/1/2, 10/5/3,  15/9/4,  24/3/5, 27/2/6, 29/16/7, 45/3/8, 48/5/9,  53/2/10,  55/5/11,  60/2/12, 62/7/13,  69/7/14, 76/6/15, 82/16/16}{
        \pgfmathtruncatemacro{\j}{mod(\c,2)}
            \edef\temp{
                \noexpand\draw[dashed, black, thick, <->] (axis cs:\x+0.1,1.2) -- (axis cs:\x+\t-0.1, 1.2);
                \noexpand\node at (axis cs:\x+\t*0.5, 1.25+0.05*\j) {$\tau_{\c}$};
            }\temp
        }
        \end{axis}
        \begin{axis}[
        yshift=-1cm,
        xlabel={$t$},
        xmin=0, xmax=100,
        width=\textwidth,
        hide y axis,
        axis lines=left,
        xlabel style={at={(axis description cs:1.02,0)}, anchor=west},
        ]
        \addplot[draw=none]{1};
        \end{axis}
        \end{tikzpicture}
    \end{subfigure}%
    \\[10pt]
    \begin{subfigure}{0.9\textwidth}
        \centering
        \begin{tikzpicture}
        \begin{axis}[
        axis lines=left,
        xlabel={$i$},
        ylabel={values $x_t$},
        xlabel style={at={(axis description cs:1.02,0)}, anchor=west},
        ylabel style={at={(axis description cs:0,1.0)}, anchor=south, rotate=-90},
        ymin=0, ymax=1,
        xmin=0, xmax=100,
        xtick={4,5,11,15,16,29,30,53,54,55,56,62,63,76,77,82,83,98,99},
        xticklabels={1,2,3,4,5,6,7,8,9,10,11,12,13,14,15,16,17,18,19},
        xticklabel style={
        yshift={-mod(\ticknum,2)*1em},
        },
        ytick={0.55},
        yticklabels={\color{c_maroon} threshold $u$},
        axis line style={thick},
        ytick style={draw=none},
        width=\textwidth,
        height=7cm,
        clip=false
        ]
        \foreach \x/\y in {1/0.5253198714088679, 2/0.23341971281509916, 3/0.017617344749315738, 4/0.6681567610447261, 5/0.6681567610447261, 6/0.4471605982116483, 
        7/0.199724013003651, 8/0.06625506773533595, 9/0.16211095998560843, 10/0.2596288226017181, 11/0.6450874534691479, 12/0.4916658748897756, 13/0.5083736737284791, 14/0.34993950033952503, 15/0.7506968652653274, 16/0.7506968652653274, 17/0.23529026261905905, 18/0.04637767363754076, 19/0.33757609097343083, 20/0.03039049735847787, 21/0.10267614778472312, 22/0.286530917552885, 23/0.008590545704922055, 24/0.2596288226017181, 25/0.5083736737284791, 26/0.17594416123117818, 27/0.24661546810190452, 28/0.0043435301166758365, 29/0.9813158095201042, 30/0.9813158095201042, 31/0.34939292907330616, 32/0.07981076336229759, 33/0.47032145301342176, 34/0.3348003990530816, 35/0.5132599825254373, 36/0.31147347978292794, 37/0.19217722749143457, 38/0.0224709405135961, 39/0.31853973551858544, 40/0.38142987818881957, 41/0.45166883916048334, 42/0.48829235895416206, 43/0.11223368717111575, 44/0.5477149605604735, 45/0.017617344749315738, 46/0.2176254699271167, 47/0.43399584725175944, 48/0.22657556753006747, 49/0.34939292907330616, 50/0.24139514252725902, 51/0.2176254699271167, 52/0.4536612047154135, 53/0.9125939032254404, 54/0.9125939032254404, 55/0.9903099777882203, 56/0.9903099777882203, 57/0.0006608249605391025, 58/0.3155178633779054, 59/0.33569455314146157, 60/0.024437017874249123, 61/0.33757609097343083, 62/0.5748709992295346, 63/0.5748709992295346, 64/0.4750296445147848, 65/0.3984607975602402, 66/0.49226491497782576, 67/0.5186879662926261, 68/0.0043435301166758365, 69/0.5083736737284791, 70/0.03545393710604023, 71/0.0454745540918483, 72/0.07775215091481588, 73/0.2699880145771954, 74/0.03303480230714859, 75/0.3954311604860138, 76/0.5973998012683518, 77/0.5973998012683518, 78/0.03545393710604023, 79/0.247446825103616, 80/0.2596288226017181, 81/0.16211095998560843, 82/0.928261484705258, 83/0.928261484705258, 84/0.5370841616889357, 85/0.24661546810190452, 86/0.08016591315093055, 87/0.5321830720090855, 88/0.11525952378724046, 89/0.1467982681790443, 90/0.22657556753006747, 91/0.2831777234416628, 92/0.13623503493324626, 93/0.16410078447327217, 94/0.004634811554663521, 95/0.30739731528253506, 96/0.2629491673958458, 97/0.22848666035997983, 98/0.7907698693079397, 99/0.7907698693079397, 100/0.408712868095583}{
            \edef\temp{
            \noexpand\draw[c_teal, thick] (axis cs:\x, 0) -- (axis cs:\x, \y);
            \noexpand\node[draw=c_teal, fill=c_teal, circle, inner sep=1pt] at (axis cs:\x,\y) {};
            }\temp
        }
        \addplot[dashed, c_maroon, thick, domain=0:12] coordinates {(0, 0.55) (100, 0.55)};
        \foreach \x/\t/\c in {4/1/1, 5/6/2, 11/4/3, 15/1/4, 16/13/5, 29/1/6, 30/23/7, 53/1/8, 54/1/9, 55/1/10, 56/6/11, 62/1/12, 63/13/13, 76/1/14, 77/5/15, 82/1/16, 83/15/17, 98/1/18}{
        \pgfmathtruncatemacro{\j}{mod(\c,2)}
            \edef\temp{
                \noexpand\draw[dashed, black, thick, <->] (axis cs:\x+0.1,1.2) -- (axis cs:\x+\t-0.1, 1.2);
                \noexpand\node at (axis cs:\x+\t*0.5, 1.25+0.05*\j) {$\tau_{\c}$};
            }\temp
        }
        \end{axis}
        \begin{axis}[
        yshift=-1cm,
        xlabel={$t$},
        xmin=0, xmax=100,
        width=\textwidth,
        hide y axis,
        axis lines=left,
        xlabel style={at={(axis description cs:1.02,0)}, anchor=west},
        ]
        \addplot[draw=none]{1};
        \end{axis}
        \end{tikzpicture}
    \end{subfigure}%
    \caption{Schematic depiction of clustering of peaks over the threshold. Top: Independent, unclustered exceedances. $\hat \Theta = 1$. Bottom: Exceedances from the time series above, now artificially modified to occur in clusters. $\hat \Theta = 0.68$.}
    \label{fig:ExtremalIndex}
\end{figure}
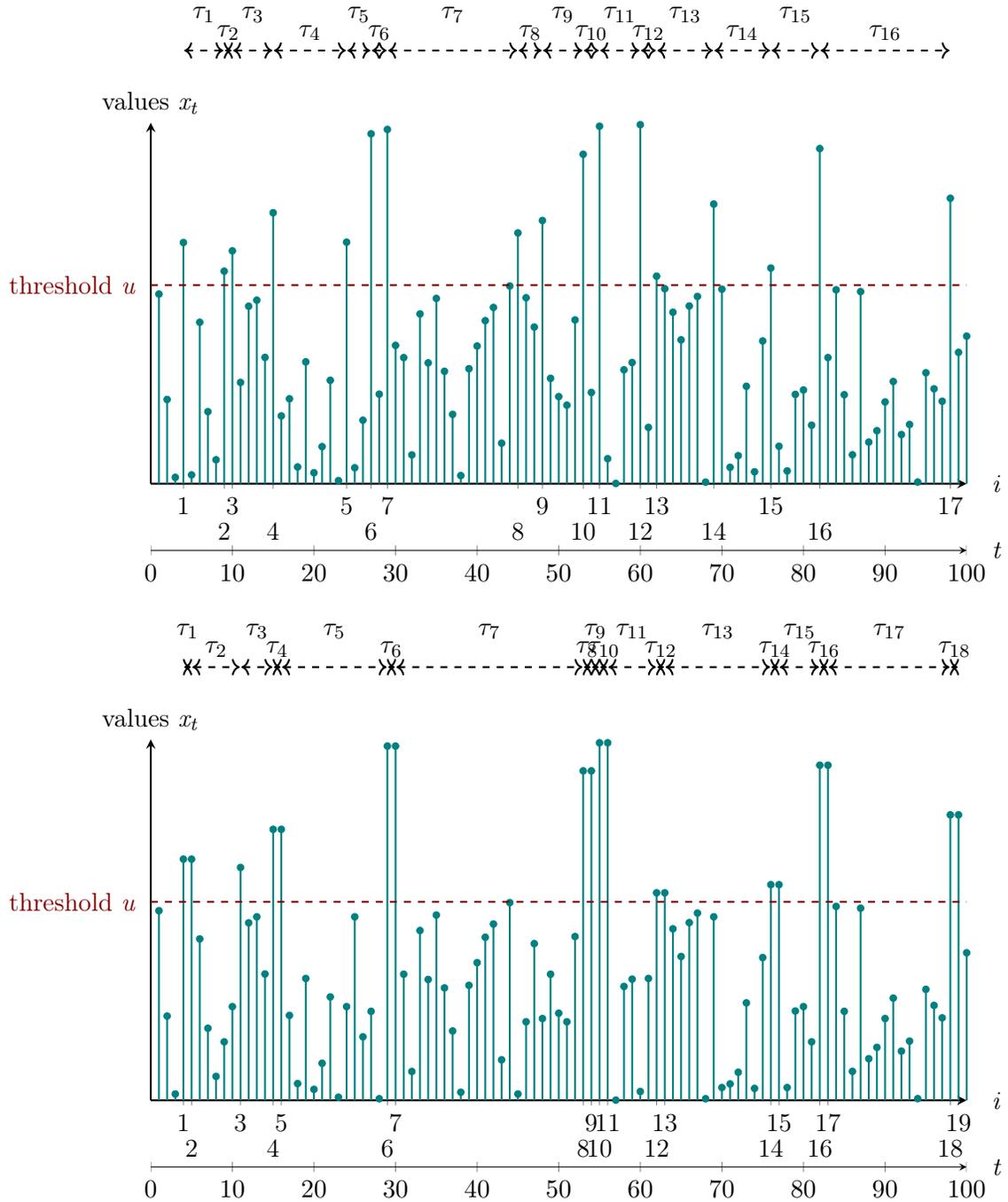

For the estimation of the mean inverse cluster size $\Theta$ we use the method described by Ferro and Segers \cite{FerroSegers2003}. The estimator $\hat \Theta$ is related to the quotient of the square of the first moment and the second moment of the empirical distribution of the $\tau_i$ according to
\begin{align}
    \hat \Theta = \begin{cases} \displaystyle \min\qty(\frac{2\langle (\tau_i-1)\rangle_i^2}{\langle (\tau_i - 1)(\tau_i - 2)\rangle_i}\,, 1), & \max \tau_i >2 \\[10pt]
    \displaystyle \min\qty(\frac{2\langle (\tau_i)\rangle_i^2}{\langle \tau_i^2\rangle_i}\,, 1), & \max \tau_i \leq 2
    \end{cases}\,, \label{eq:cluster_estimator}
\end{align}
where $\langle\, \cdot\, \rangle_i$ indicates an average over all $n-1$ time spans $\tau_i$. This is the estimator version of Eq.~\eqref{eq:ferro_segers_theory}, which we motivated already in Sec.~\ref{ss:extremalindex}. In the case of a perfect Poisson process $\hat \Theta$ evaluates to 1, while for clustered exceedances it is smaller and scales with the real mean inverse cluster size $\Theta$ of the process. The smaller $\Theta$, the larger is the average cluster of exceedances generated by the process. This is schematically depicted in Fig.~\ref{fig:ExtremalIndex}, where two illustrative examples are shown. In the upper panel, where peaks are generated independently, the estimator in Eq.\eqref{eq:cluster_estimator} yields $\hat \Theta = 1$. In the lower panel, the same peaks are artificially modified to occur in clusters. For that, some values following a peak exceeding the threshold are augmented to also exceed the threshold. Others are lowered to fall below the threshold, to keep the number of exceedances comparable. This leads to a reduced value, $\hat \Theta \simeq 0.68$.
\par
\begin{figure}
    \centering
    \begin{tikzpicture}
    \begin{axis}[
    domain=0.5:1.2,
    samples=200,
    axis lines=left,
    xlabel={value $x$},
    xlabel style={at={(axis description cs:1.02,0)}, anchor=west},
    ylabel style={at={(axis description cs:0,1.05)}, anchor=south, rotate=-90},
    ymin=0, ymax=1,
    xmin=0, xmax=7.2,
    xtick={5},
    xticklabels={\color{c_maroon}threshold $u$},
    ytick={},
    extra y ticks={0.9184}, 
    extra y tick labels={\color{c_maroon} quantile $1-\alpha$},
    axis line style={thick},
    tick style={thick},
    width=12cm,
    height=8cm,
    clip=false
    ]
    
    \addplot[c_teal, domain=0:7, samples=100, only marks,
    mark=*,
    mark options={scale=0.4}, ybar interval, area style, mark=no, opacity=0.3] ({x}, {1-(1+0.5*x)^(-1/0.5)});
    
    \addplot[dashed, c_maroon, thick, domain=0:7] coordinates {(5, 0) (5, 1)};
    \addplot[dashed, c_maroon, thick, domain=0:7] coordinates {(0, 0.9184) (5, 0.9184)};
    
    \draw[dashed, c_maroon, thick, <-] (axis cs:4.5,0.5) -- (axis cs:5, 0.5);
    \draw[dashed, c_maroon, thick, ->] (axis cs:5,0.5) -- (axis cs:5.5, 0.5);
    
    \addplot[line width=3pt, c_orange, domain=5:7, samples=100, label={GPD fit}]
        ({x}, {1-(1+0.5*x)^(-1/0.5)});
    
    \draw[decorate, decoration={brace, mirror, amplitude=8pt}, thick]
        (axis cs:0,0) -- (axis cs:5,0) node[midway, yshift=-0.8cm] {not extreme};
    
    \draw[decorate, decoration={brace, mirror, amplitude=8pt}, thick]
        (axis cs:5,0) -- (axis cs:7,0) node[midway, yshift=-0.8cm] {extreme};
    
    \draw[->, thick, c_orange]  (axis cs:6.5,1.05) -- (axis cs:6.4,1.0);
    \node[c_orange] at (axis cs:6.5,1.1) {GPD fit (tail)};
    
    \draw[->, thick, c_teal]  (axis cs:1.5,0.95) -- (axis cs:1.9,0.8);
    \node[c_teal] at (axis cs:1.5,1) {stylized empirical CDF};
    
    \end{axis}
    \end{tikzpicture}
    \caption{Schematic depiction of the estimation procedure.}
    \label{fig:SchematicFitting}
\end{figure}
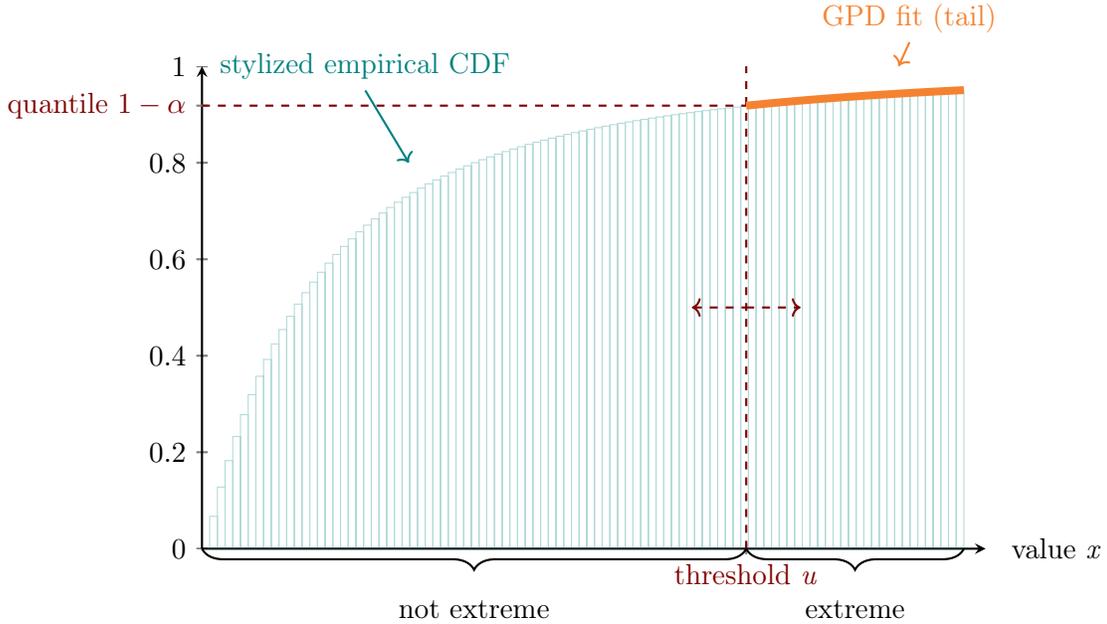

 To evaluate the tail behavior we use Maximum-Likelihood estimation for the parameters of the generalized Pareto distribution \eqref{eq:GPD} of the peaks excessing a certain threshold. Figure~\ref{fig:SchematicFitting} shows a schematic depiction of the procedure. The log-likelihood of $G^{\text{GPD}}_{\gamma,\sigma}$, reads
\begin{align}
    l(\gamma,\sigma) = \begin{cases} \displaystyle - n \ln \sigma - \qty(1+\frac{1}{\gamma})\sum_{i=1}^n \ln(1+\gamma\frac{x_i}{\sigma}), & \text{ for } \gamma \neq 0\\ 
    \displaystyle -n \ln \sigma  - \sum_{i=1}^n \frac{x_i}{\sigma}, & \text{ for } \gamma = 0
    \end{cases}\,, \label{eq:loglikelihood}
\end{align}
this function is maximized to obtain $\hat \gamma$ and $\hat \sigma$.
From Eq.~\eqref{eq:loglikelihood}, asymptotic variances as the squared standard errors of $\hat \gamma$ and $\hat \sigma$ are obtained in the context of the Fisher information matrix by Smith \cite{Smith1984} as
\begin{align}
    \operatorname{var}(\hat \gamma) \sim \frac{1}{n}(1+\gamma)^2, && \operatorname{var}(\hat \sigma) \sim \frac{1}{n}2\sigma^2(1+\gamma)\, . \label{eq:asymptoticvariances}
\end{align}
They are accessibly provided in Ref.~\cite{Hosking1987}. To evaluate the bias or goodness-of-fit we use the normalized root mean square deviation (NRMSD) related to quantile-quantile (Q-Q) plots. These are plots of the sorted time series $\qty{x_i}_{i=1, \ldots, n}$ of peaks over threshold $u$ against ${G^{\text{GPD}}_{\hat\gamma, \hat\sigma}}^{-1}\qty(F_{u}^{\text{empirical}}(x_i))$, so we plot the empirical observations against the theoretically implied quantiles. If the estimated distribution were a perfect fit, we should see a straight line. The NRMSD as deviation from the straight line is calculated as
\begin{align}
    \rho = \frac{1}{\max_i(x_i) - \min_i(x_i)} \sqrt{\frac{1}{n} \sum_{i=1}^n \qty({G^{\text{GPD}}_{\hat\gamma, \hat\sigma}}^{-1}\qty(F_{u}^{\text{empirical}}(x_i)) - x_i)^2}\,. \label{eq:NRMSD}
\end{align}
Due to the bias-variance trade-off, the expected behavior of $\rho$ is nonmonotonic. As the threshold $u$ increases, the fit of the generalized Pareto distribution improves because the model becomes more appropriate for increasingly extreme data, leading to a decrease in $\rho$. However, beyond a certain threshold, the number of exceedances becomes too small which increases estimation variance and amplifies the influence of individual extreme observations. Therefore the fit loses quality and $\rho$ increases again.

\FloatBarrier

\section{Application to Multivariate Setting \label{s:multivariate_setting}}
Section~\ref{ss:rotated_returns} addresses the second point of the paper's guiding questions: rotation into the eigenbasis of the correlation matrix. In Sec.~\ref{ss:portfolio_interpretation} we shortly present the connection to portfolio risk management.

\subsection{Rotated Returns \label{ss:rotated_returns}}
\begin{figure}[t]
    \centering
    \begin{subfigure}{0.5\textwidth}
        \includegraphics[width=\textwidth]{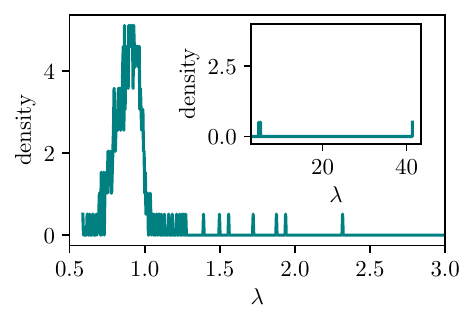}
    \end{subfigure}%
    \begin{subfigure}{0.5\textwidth}
        \includegraphics[width=\textwidth]{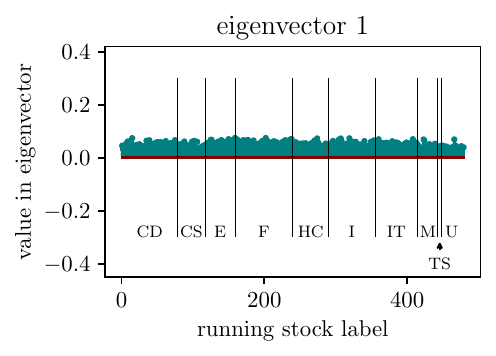}
    \end{subfigure}\\
    \begin{subfigure}{0.5\textwidth}
        \includegraphics[width=\textwidth]{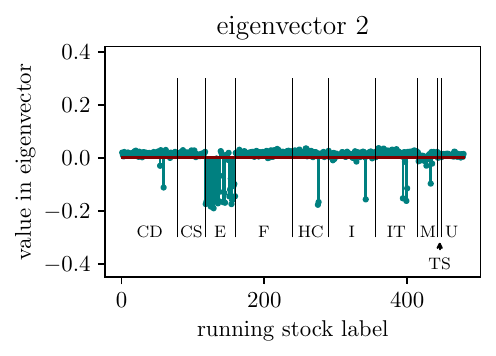}
    \end{subfigure}%
    \begin{subfigure}{0.5\textwidth}
        \includegraphics[width=\textwidth]{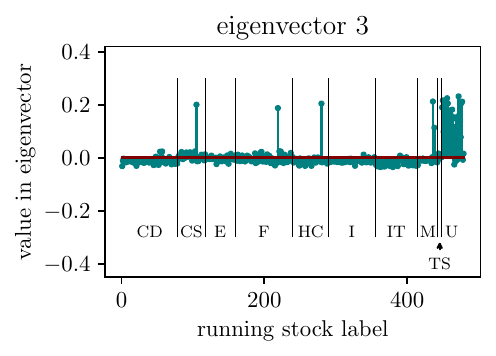}
    \end{subfigure}
    \caption{Top left: Spectral density of the financial correlation matrix ($\Delta t = 1$s) of 2014. Top right and bottom: Largest three normalized eigenvectors visibly corresponding to the market, the Energy sector and the Utility sector of the GICS \cite{GICS}.}
    \label{fig:eigenvalue_density}
\end{figure}

From the normalized data matrix we calculate a correlation matrix, with the Pearson correlation coefficient for each pair of stocks as its elements,
\begin{align}
    C = \frac{1}{T} MM^\dagger\, . \label{eq:CorrMatrix}
\end{align}
Since $C$ is a symmetric, positive semidefinite matrix, the eigenvalues $\Lambda_k, k=1, \ldots, K$, are real and the diagonalizing matrices $U, U^\dagger$ obtained from the spectral decomposition
\begin{align}
    C = U\Lambda U^\dagger\,, \qqtext{ where }\Lambda = \operatorname{diag}(\Lambda_1, \ldots, \Lambda_K)\,, \label{eq:SpectralDecomposition}
\end{align}
are orthogonal. In our case $C$ is positive definite. Random matrix theory suggests to interpret the large bulk in the eigenvalue density shown in Fig.~\ref{fig:eigenvalue_density} as random noise, while it is well known that the largest eigenvalue and its eigenvector encode information on the market as a whole \cite{Laloux1999}.
The few smaller eigenvalues outside the bulk mainly correspond to industrial sectors, see Ref.~\cite{Plerou2000} and Fig.~\ref{fig:eigenvalue_density} in the present contribution. We calculate the rotated and rescaled returns as
\begin{align}\label{eq:rotated_rescaled}
    R = \Lambda^{-1/2}U^\dagger M\, ,
\end{align}
 where we choose the positive sign when calculating the square root. By definition the $\sqrt{\Lambda_k}$ are the standard deviations of the rotated time series $(U^\dagger M)_k$, $k=1, \ldots, K$, thus $R$ contains indeed rescaled and thus comparable time series $R_k$ as its rows. In the statistics literature, this transformation \eqref{eq:rotated_rescaled} is usually called PCA--whitening \cite{KessyWhitening2018}, while in climate research this and similar approaches are often referred to as empirically orthogonal function (EOF) analyses. In climate research, these approaches are extensively used for reducing spatial dimensionality, see Ref.~\cite{PhillipsKantz2025} as a typical and recent application. We refer to these rotated and rescaled time series as modes and number them according to the size of the eigenvalues, starting with the largest. The first thus reflects market behavior, while the second mode corresponds to the sector with the highest variance and so forth.
 
\begin{figure}[t]
    \centering
    \begin{subfigure}{0.5\textwidth}
        \includegraphics[width=\textwidth]{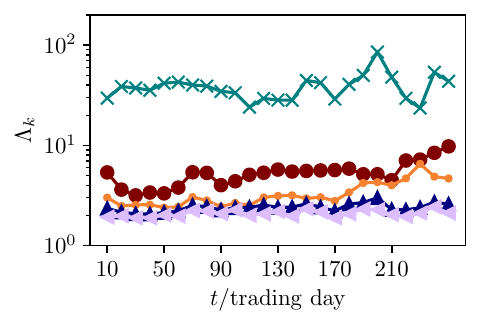}
    \end{subfigure}%
    \begin{subfigure}{0.5\textwidth}
        \includegraphics[width=\textwidth]{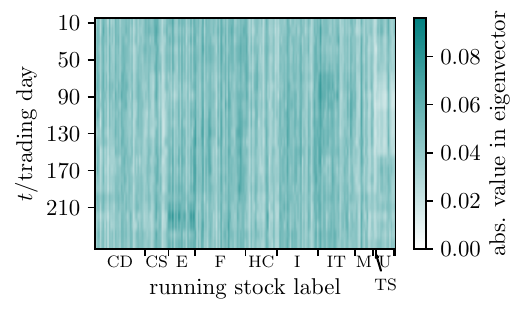}
    \end{subfigure}\\
    \begin{subfigure}{0.5\textwidth}
        \includegraphics[width=\textwidth]{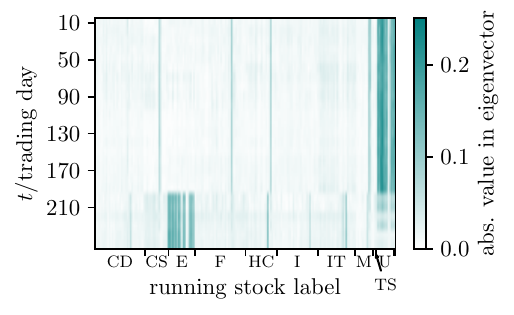}
    \end{subfigure}%
    \begin{subfigure}{0.5\textwidth}
        \includegraphics[width=\textwidth]{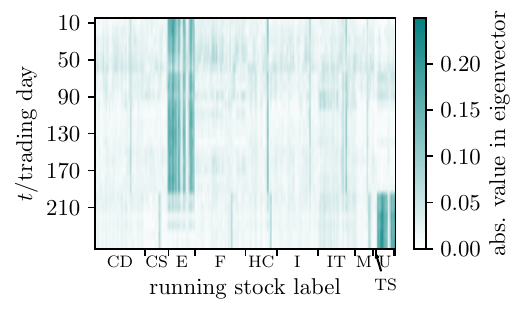}
    \end{subfigure}
    \caption{Top left: Largest five eigenvalues of the financial correlation matrices ($\Delta t = 1$s) evaluated in disjoint windows of 10 days throughout the year 2014. Top right and bottom: Absolute values of the entries for the largest three normalized eigenvectors evaluated in disjoint windows of 10 days throughout the year 2014.}
    \label{fig:eigenvalue_evolution_in_time}
\end{figure}

In general one cannot view the eigenvalues as fixed in time and also the composition and therefore the interpretation of the eigenvectors is subject to changes in a nonstationary system. In an exemplary analysis we see such nonstationarities if we calculate the correlation matrix and analyze its spectrum in disjoint windows of 10 trading days. The results are shown in Fig.~\ref{fig:eigenvalue_evolution_in_time}, for the same eigenvectors as plotted in Fig.~\ref{fig:eigenvalue_density}.
Importantly, we cannot compare the results directly to our average correlation matrix $C$, since the time range used for calculation of the correlation matrix in each window is much smaller, but we see indeed similar interpretations for the eigenvectors as market and sectors arising. Although two eigenvalues cannot cross each other, when they come close their corresponding eigenvectors may ``flip'' or, in other words, swap their meaning. This phenomenon is visible for the second and third eigenvector in Fig.~\ref{fig:eigenvalue_evolution_in_time}. It is also possible that the interpretation changes rather continuously over time. However, in this paper we restrict ourselves to a set of fixed collective coordinates, arising from the fixed $C$ as an average over the whole year 2014. Thus in the following the term ``nonstationarity'' will not be used in a way including the nonstationarity of the definition of collective coordinates.
 
At small return horizons, $\Delta t$ of $1$s or comparable, the measurement of the correlation coefficients is influenced by the so-called Epps effect \cite{Epps1979}, and especially by tick-size discretization and asynchronicity effects \cite{HayashiYoshida2005, Münnix2010, MünnixTickSize2010}. Due to these effects, the correlation coefficients are decreasing with decreasing return horizon. Thus, when rotating into the eigenbasis \eqref{eq:rotated_rescaled}, the Epps effect carries over to the multivariate analysis of returns. In the present work we nevertheless deliberately focus on
$\Delta t=1$s returns. Our objective is not to reconstruct the latent efficient-price dynamics or the correlation matrix that would be obtained after removing all microstructure effects. Rather, we analyze the extreme value statistics of the empirically observed high-frequency returns themselves. The horizon of $1$s provides a sufficiently large number of observations for a statistically meaningful investigation of the tails, while still retaining the high-frequency features that are lost under aggregation to daily returns. Although the Epps effect reduces the numerical values of the correlations, the correlation matrix at $\Delta t=1$s still exhibits meaningful collective structure, in particular through its leading eigenvectors, as we showed in Fig.~\ref{fig:eigenvalue_density}. While the effect does not uniformly attenuate correlations across all asset pairings, our reliance on linear combinations, which imply at least some averaging, ensures that such pairwise heterogeneity is unlikely to substantially bias the aggregate results.
The effect of tick size discretization is not only visible in the drop in correlation values, but also affects the marginal distributions of individual return time series, and thus an analysis of extreme value statistics at high frequencies. However the tick size affects different stocks differently. Thus due to the rotation, which is just a linear combination of return time series, the tick size effect is washed out, and the distributions appear smoothed, allowing for an extreme value analysis.

\begin{figure}
    \centering
    \includegraphics{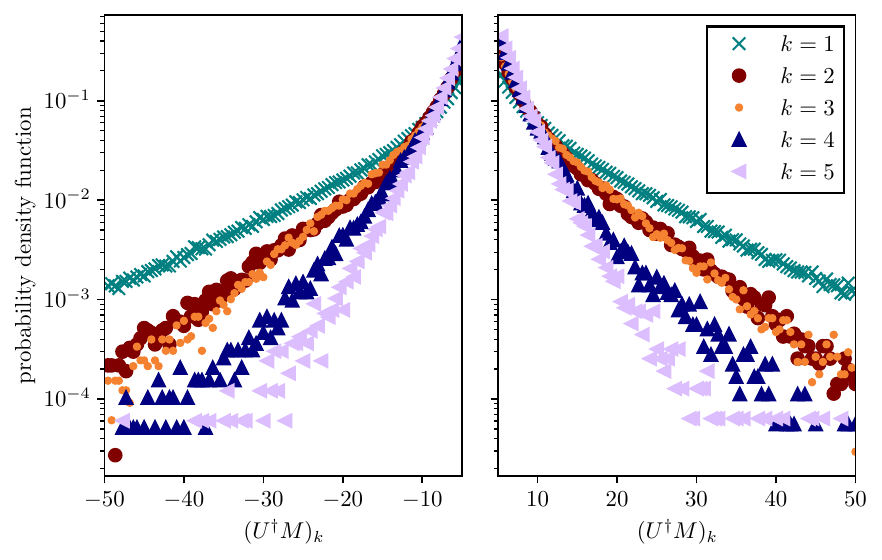}
    \caption{Left and right tails of the probability density functions for rotated returns referring to the five largest eigenvalues, without the rescaling $1/\sqrt{\Lambda_k}$.}
    \label{fig:distributions_unscaled}
\end{figure}

\begin{figure}
    \centering
    \includegraphics{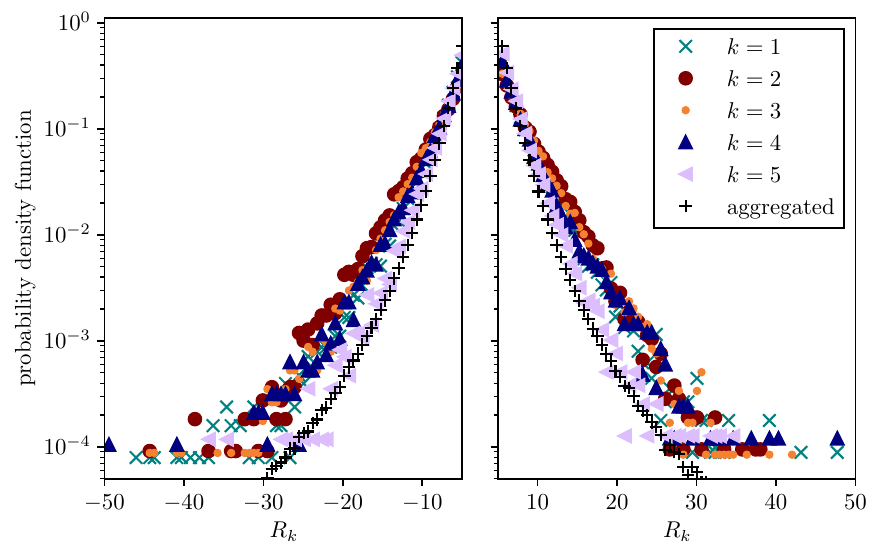}
    \caption{Left and right tails of the probability density functions for the modes (rotated and rescaled) referring to the five largest eigenvalues and the aggregation of all modes.}
    \label{fig:distributions_rescaled}
\end{figure}

\begin{figure}
    \centering
    \begin{subfigure}{0.5\textwidth}
    \begin{overpic}[width=\textwidth]{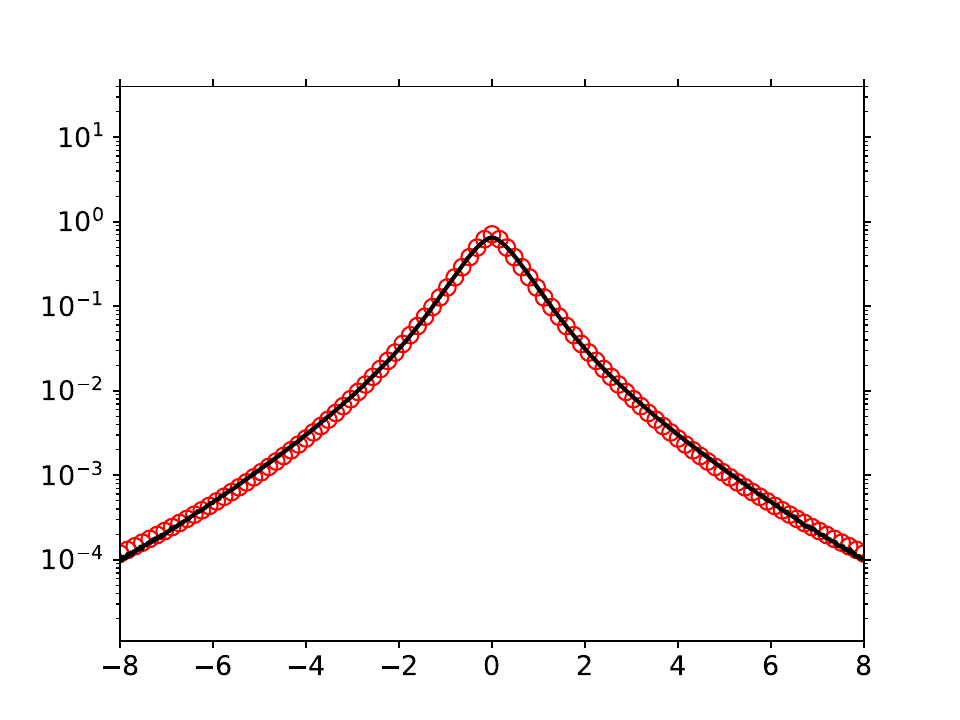}
				\put(18,55){\noindent\fbox{\parbox{1.7cm}{50 days\\$\Delta t = 1\,s$}}}
				\put(50,0){\makebox(0,0){aggregated return}}
				\put(1,35){\makebox(0,0){\rotatebox{90}{pdf}}}
	\end{overpic}
	\end{subfigure}%
	\begin{subfigure}{0.5\textwidth}
    \begin{overpic}[width=\textwidth]{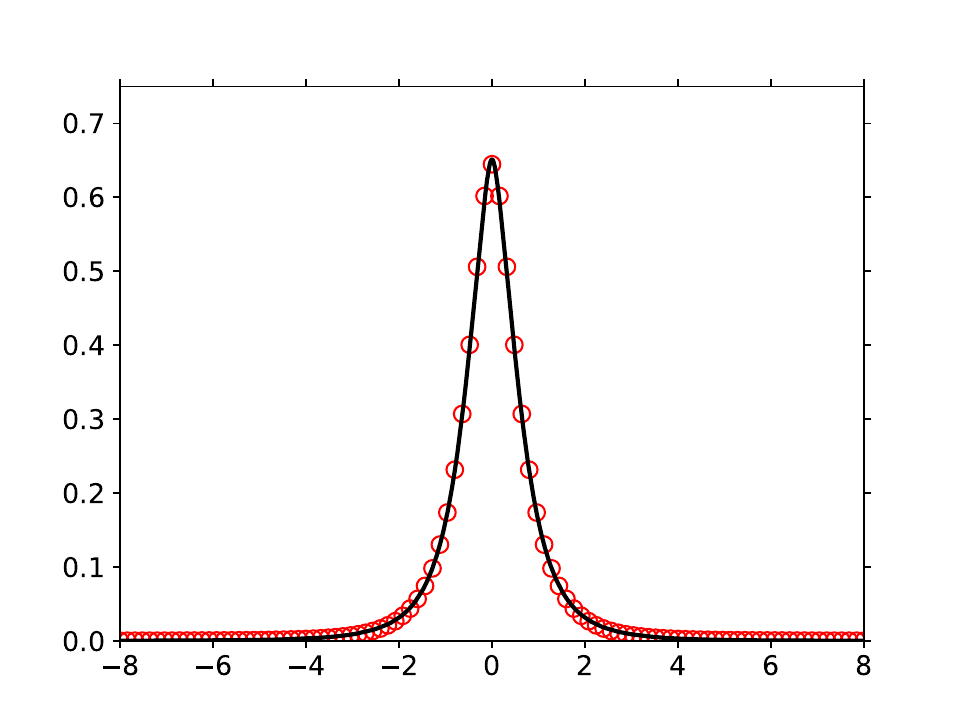}
				\put(18,55){\noindent\fbox{\parbox{1.7cm}{50 days\\$\Delta t = 1\,s$}}}
				\put(50,0){\makebox(0,0){aggregated return}}
				\put(1,35){\makebox(0,0){\rotatebox{90}{pdf}}}
			\end{overpic}
	\end{subfigure}%
    \caption{Probability density function for aggregated returns in a time interval of 50 days in 2014, with return horizon 1s. Obtained empirically (black line) and generated by a random matrix model (red circles). Left: Ordinate logarithmically scaled. Right: Ordinate linearly scaled. Adapted from Ref.~\cite{Heckens2025_II}.}
    \label{fig:RMTpaperII}
\end{figure}

Regarding an in-depth analysis and a random matrix model for the multivariate distributions of the modes, we refer to Refs.~\cite{Manolakis_I, Heckens2025_II}. For the convenience of the reader, we briefly sketch relevant details here shortly. We see the influence of the rescaling $\Lambda^{-1/2}$ from Eq.~\eqref{eq:rotated_rescaled} in Figs.~\ref{fig:distributions_unscaled} and \ref{fig:distributions_rescaled}, where the tails of the probability densities for the rotated returns are plotted without and with the rescaling, respectively.%
In Fig.~\ref{fig:distributions_rescaled} we also plotted the aggregated probability density function, \textit{i.e.} the density which emerges if we lump all $R_k$ together. As explained in Ref.~\cite{Heckens2025_II}, this represents the average density referring to an eigenvalue of the bulk. This density was reproduced in Ref.~\cite{Heckens2025_II} by a random matrix model, see Fig.~\ref{fig:RMTpaperII}.%
Figure~\ref{fig:distributions_unscaled} shows that the unscaled densities are ordered in accordance to the magnitude of the eigenvalues, while the rescaling in Fig.~\ref{fig:distributions_rescaled} removes this variance ordering. The densities are still not exactly the same, however the reason now is not the variance but the pure tail behavior, thus the ordering is now changed. Interestingly, in the range relevant for extreme values of the first five modes $5 \lesssim \abs{R_k} \lesssim 20$, we notice that the aggregated density mostly holds less weight than the density of the modes. For larger $\abs{R_k} \gtrsim 20$ a comparison between densities is difficult due to the drastic difference in statistical significance for aggregated and single $k$. However, this means that the first few modes referring to the collective market and sectors show greater risk than the average mode referring to an eigenvalue from the bulk and thus a structureless, random linear combination of returns. 

We have defined the rotation and rescaling \eqref{eq:rotated_rescaled} that transforms correlated returns into mutually uncorrelated, variance-normalized modes.
In the remainder of the paper, extreme value statistics are applied to these modes rather than to the raw asset returns. In Sec.~\ref{s:dataanalysisresults} and Sec.~\ref{s:local_estimation} we analyze these first five modes with regard to their tail behavior.

\subsection{Connection to Portfolio Risk Management \label{ss:portfolio_interpretation}}
If we consider a specific mode and therefore fix $k$, we have
\begin{align}
    R_k(t) = \Lambda_k^{-1/2} \sum_{l=1}^K U_{lk} M_l(t)\, .
\end{align}
This resembles the portfolio return $W(t)$ in modern portfolio theory, which is usually defined by
\begin{align}\label{eq:portfolio_def}
    W(t) = \sum_{l=1}^K g_l(t) Y_l(t), && \sum_{l=1}^K g_l(t) = 1\,,
\end{align}
where $Y_k(t)$ is the return of asset $l$ at time $t$ and $g_l(t)$ is the fractional weight of asset $l$ in the portfolio. Thus we can view our modes $R_k(t)$ as similar to $W(t)$ for distinct choices of portfolios. The entries $U_{lk}$ of the eigenvectors take the role of the weights $g_l$ in the case of no turnover. In Fig.~\ref{fig:eigenvalue_density} we see that all stocks contribute somewhat equally to the first eigenvector corresponding to the largest eigenvalue, which therefore resembles a diversified portfolio choice covering the whole market. Thus our first mode $R_1(t)$ is similar to the return of such a market-wide portfolio. Since the portfolio is broadly diversified, it is exposed to the pure systematic risk, while idiosyncratic risk is largely diversified away. Contrasting that, the entries in the second eigenvector, corresponding to the second largest eigenvalue, are mostly constituted of stocks from the Energy sector, thus $R_2(t)$ is similar to the return of an investment portfolio focused on energy. Such a sector-concentrated portfolio is exposed to both systematic risk and substantial idiosyncratic risk, the latter of which can be reduced by broader diversification. In the following we analyze the extremal behavior of the $R_k(t)$, and we will see that the tail behavior differs. \par 
We emphasize that the present analysis is retrospective: the term risk management is used here in the sense of identifying and quantifying risk carried by empirically observed collective modes, not in the sense of providing an out-of-sample forecasting or trading procedure. An operational implementation would require estimating the correlation matrix, eigenbasis and quantities introduced later from a prior calibration window and updating them sequentially, which constitutes a natural extension of the present framework.

\FloatBarrier

\section{\label{s:dataanalysisresults} Estimation for Marginal Distribution}
The return modes differ in their tail behavior. As linear combinations of stocks with similar dynamics, they capture correlated behavior. In the following section, we apply the POT approach to the marginal data of the modes corresponding to the five largest eigenvalues. This addresses the third point of the paper's guiding questions.

\begin{figure}
    \centering
    \includegraphics{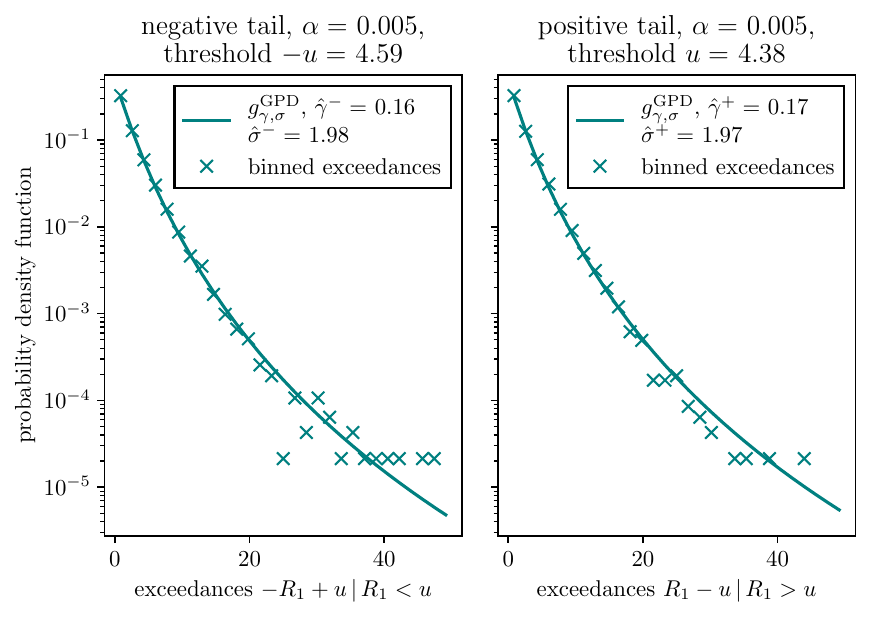}
    \caption{Binned exceedances for both tails of the first mode with tail-quantile $\alpha = 0.005$ together with the theoretical density function $g_{\gamma, \sigma}^{\text{GPD}}$ obtained from the fits.}
    \label{fig:pdf_gpd_fit}
\end{figure}

\begin{figure}
    \centering
    \begin{subfigure}{0.75\textwidth}
        \includegraphics[width=\textwidth]{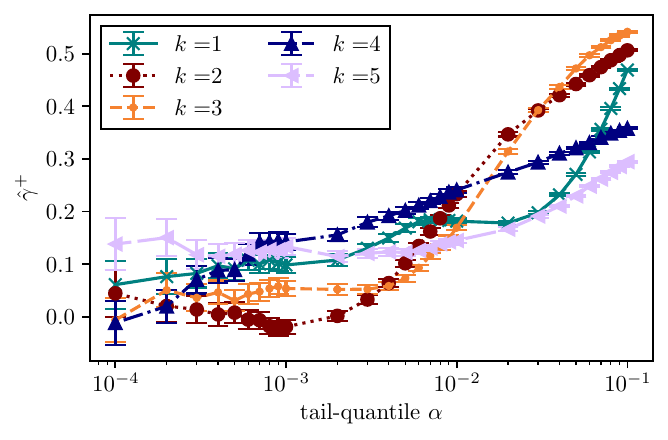}
    \end{subfigure}%
	\\%
	\begin{subfigure}{0.75\textwidth}
		\includegraphics[width=\textwidth]{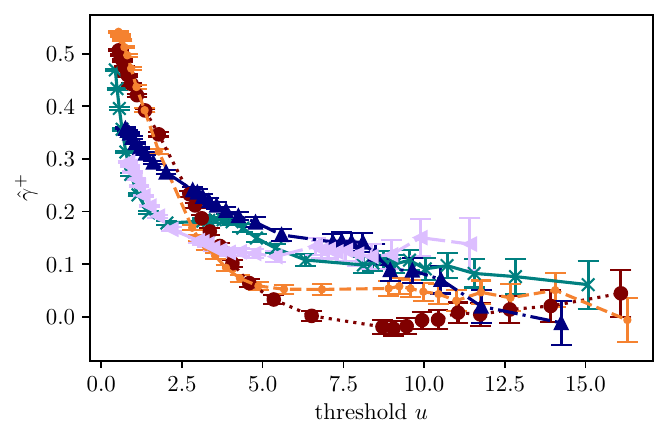}
	\end{subfigure}%
	\\%
    \begin{subfigure}{0.5\textwidth}
        \includegraphics[width=\textwidth]{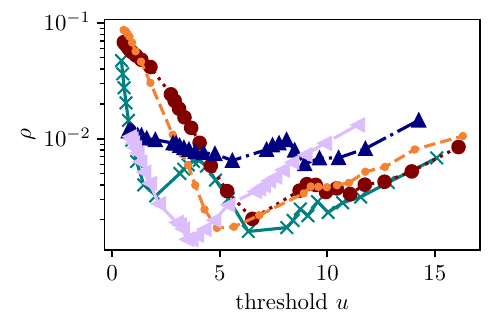}
    \end{subfigure}
    \caption{Naive estimation of tail shape parameter without declustering extreme values. Top: Estimated $\hat \gamma^+$ against tail-quantile $\alpha$. Middle: Estimated $\hat \gamma^+$ against threshold $u$. Standard errors according to \eqref{eq:asymptoticvariances}. Bottom: NRMSD of estimation.}
    \label{fig:naive_estimation+}
\end{figure}

\begin{figure}
	\centering
	\begin{subfigure}{0.75\textwidth}
		\includegraphics[width=\textwidth]{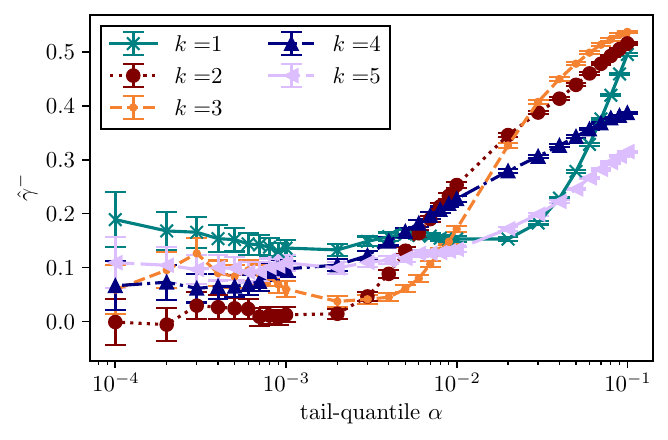}
	\end{subfigure}%
	\\%
	\begin{subfigure}{0.75\textwidth}
		\includegraphics[width=\textwidth]{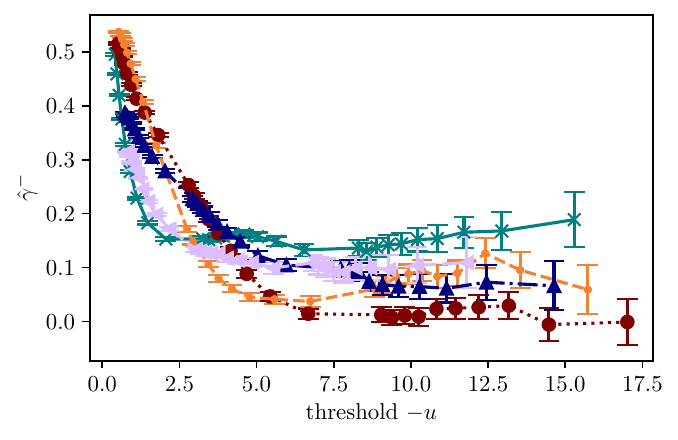}
	\end{subfigure}%
	\\%
	\begin{subfigure}{0.5\textwidth}
		\includegraphics[width=\textwidth]{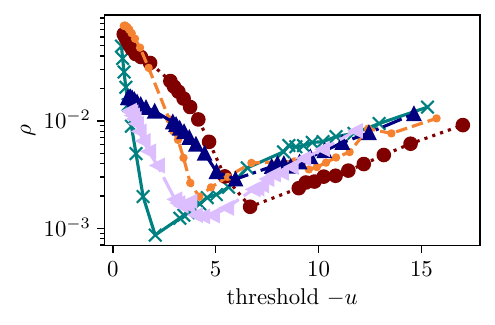}
	\end{subfigure}
	\caption{Naive estimation of tail shape parameter without declustering extreme values. Top: Estimated $\hat \gamma^-$ against tail-quantile $\alpha$. Middle: Estimated $\hat \gamma^-$ against threshold $-u$. Standard errors according to \eqref{eq:asymptoticvariances}. Bottom: NRMSD of estimation.}
	\label{fig:naive_estimation-}
\end{figure}

In theory the estimation only using the marginal distribution should suffice, if the dynamics are stationary, the threshold is large enough and the condition $\textbf{D}$ is met, as mentioned in Sec.~\ref{s:extremevaluetheory}. We denote the tail-quantile by $\alpha$, implicitly defining the threshold $u$ for estimation, as sketched in Fig.~\ref{fig:SchematicFitting}. Results for the tail behaviors $\hat \gamma^+$ (positive tail), and $\hat \gamma^-$ (negative tail) are depicted versus tail-quantile $\alpha$ and threshold $u$ in Fig.~\ref{fig:naive_estimation+}, respective Fig.~\ref{fig:naive_estimation-}. In addition the NRMSD is shown, to provide an estimate of the goodness-of-fit. To illustrate how the fitted probability densities appear, we depict an example in Fig.~\ref{fig:pdf_gpd_fit}.
In this Fig.~\ref{fig:pdf_gpd_fit}, we show the empirical density of exceedances together with the corresponding theoretical probability distribution Eq.~\eqref{eq:GPD_density}, with fit parameters determined as explained above, for an exemplary tail-quantile and the first mode.

We find that the minimum of the NRMSD coincides with the optimal threshold, which is also visible in a stabilisation of the fitted tail indices $\hat \gamma$. At higher thresholds, the standard errors of the tail indices are rather large, hence a clear ranking for all tails is possible only in an intermediary regime, around the minima of the NRMSD. For both tails and for all modes, the estimated $\hat\gamma$ differs, but the behavior is mostly of the Fréchet-type. This is expected for high-frequency financial returns which are heavy-tailed, \textit{cf.} Ref.~\cite{Plerou1999}. It is strongly dependent on the threshold $u$, similar behavior was already reported in Refs.~\cite{GopikrishnanScaling1999, Plerou1999} for individual companies and indices. However, the reported values there refer to returns with much larger return horizons $\Delta t$, starting at a scale of minutes. In addition, we see in the NRMSD that the optimal threshold for estimation differs for different modes, and that all modes converge to the GPD with different velocities.

\begin{figure}
	\centering
	\includegraphics[width=0.75\textwidth]{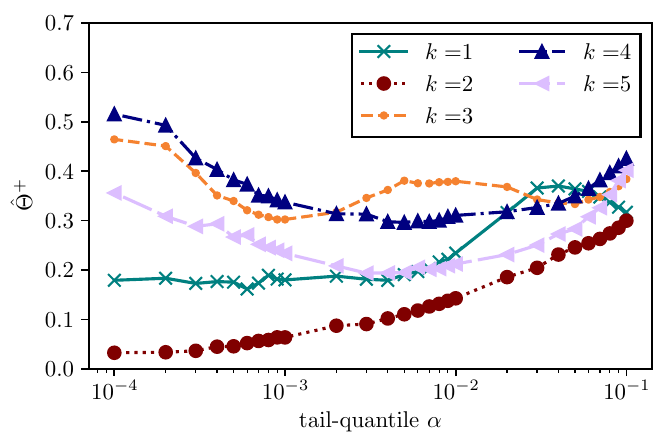}
	\caption{ Estimated extremal index $\hat \Theta^+$ against tail-quantile $\alpha$.}
	\label{fig:theta_naive+}
\end{figure}

\begin{figure}
	\centering
	\includegraphics[width=0.75\textwidth]{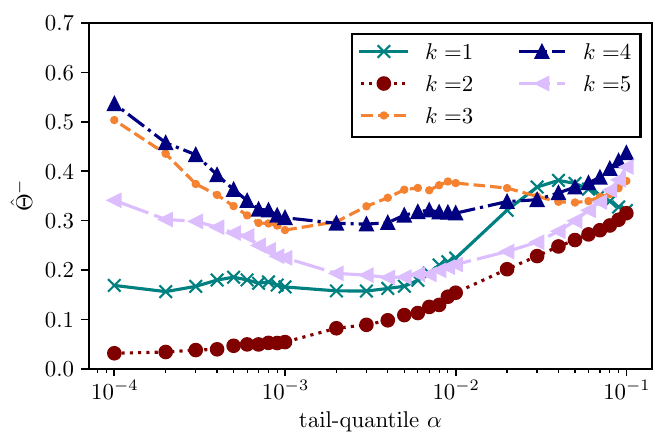}
	\caption{Estimated extremal index $\hat \Theta^-$ against tail-quantile $\alpha$.}
	\label{fig:theta_naive-}
\end{figure}

\begin{figure}
	\centering	\includegraphics[width=0.75\textwidth]{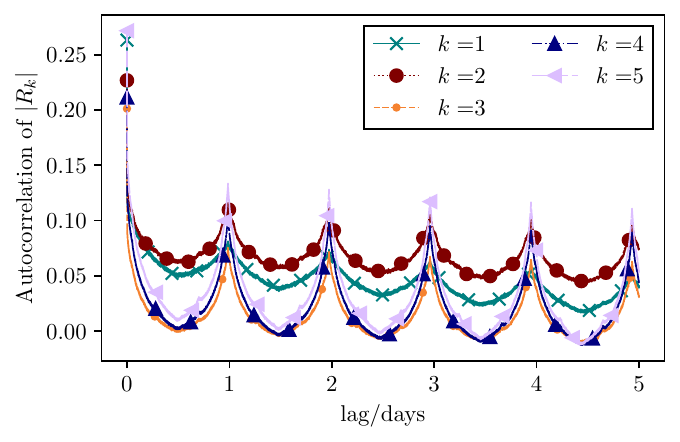}
	\caption{Autocorrelation of absolute values of rotated return modes $\vert R_k \vert$, against time lag. The first value for each mode is plotted at a lag of 1s. Markers are evenly spaced for better readability, data frequency is 1s.}
	\label{fig:autocorr_abs}
\end{figure}

The results for the clustering indices $\hat \Theta^+$ (positive tail) and $\hat \Theta^-$ (negative tail) are shown in Figs.~\ref{fig:theta_naive+} and \ref{fig:theta_naive-}. 
The estimate of the extremal index $\hat \Theta$ differs from unity in all cases, indicating clustered extremes in all evaluated modes and for both tails. This is consistent with volatility clustering, which is reflected in the clustering of extreme values. The effect is not limited to small tail-quantiles $\alpha$, which implies that even at longer time scales there are correlations between extremes. The fact that financial markets show autocorrelations in nonlinear transformations of returns on a long timescale, is well-established \cite{Ding1993}. In Fig.~\ref{fig:autocorr_abs}, the autocorrelation for the absolute values of returns is plotted for the five first modes.
We notice that the different levels of persistence in the autocorrelations of the modes are in accordance with the picture from our analysis of the extremal index. \par 
In Tab.~\ref{tab:marginal_est_mean_pos} we show the mean values of $\hat \gamma^+$ and $\hat \Theta^+$ evaluated from the fits corresponding to the ten lowest values of the NRMSD. In addition, we also show the standard deviations. Table~\ref{tab:marginal_est_mean_neg} shows the same for the negative tails, \textit{i.e.} for $\hat \gamma^-$ and $\hat \Theta^-$.
\begin{table}[htb]
\caption{Mean values and standard deviations of $\hat \gamma^+$ and $\hat \Theta^+$ evaluated from the fits corresponding to the ten lowest values of the NRMSD.}
\label{tab:marginal_est_mean_pos}
\centering
\begin{tabular}{@{}lllll@{}}
\toprule
mode & $\langle \hat \gamma^+ \rangle$ & std. $\hat \gamma^+$ & $\langle \hat \Theta^+ \rangle$ & std. $\hat \Theta^+$ \\ \midrule
1    & 0.126                      & 0.030                    & 0.183                      & 0.010                    \\
2    & 0.026                      & 0.053                    & 0.080                      & 0.023                    \\
3    & 0.044                      & 0.019                    & 0.337                      & 0.046                    \\
4    & 0.111                      & 0.041                    & 0.377                      & 0.050                    \\
5    & 0.125                      & 0.005                    & 0.219                      & 0.026                    \\ \bottomrule
\end{tabular}%
\end{table}%
\begin{table}[htb]
\caption{Mean values and standard deviations of $\hat \gamma^-$ and $\hat \Theta^-$ evaluated from the fits corresponding to the ten lowest values of the NRMSD.}
\label{tab:marginal_est_mean_neg}
\centering
\begin{tabular}{@{}lllll@{}}
\toprule
mode & $\langle \hat \gamma^- \rangle$ & std. $\hat \gamma^-$ & $\langle \hat \Theta^- \rangle$ & std. $\hat \Theta^-$ \\ \midrule
1    & 0.158                      & 0.009                    & 0.220                      & 0.067                    \\
2    & 0.020                      & 0.011                    & 0.055                      & 0.016                    \\
3    & 0.066                      & 0.019                    & 0.317                      & 0.030                    \\
4    & 0.094                      & 0.026                    & 0.324                      & 0.031                    \\
5    & 0.119                      & 0.011                    & 0.198                      & 0.012                    \\ \bottomrule
\end{tabular}%
\end{table}%
We see that the second mode, corresponding to the Energy sector, is generally the one with the smallest tail index, if we consider thresholds where the standard error is rather small and the NRMSD is close to its minimum. The positive tail even crosses into the Gumbel or Weibull domain for some thresholds in this range. In addition, it is also the mode with strongest serial clustering and persistence of volatility autocorrelations. These effects are even stronger than the effect on the first mode, \textit{i.e.}, the whole market itself, and suggest that the Energy sector exhibits a different extremal behavior than the market and other sectors in the year 2014.

\begin{figure}
    \centering
    \includegraphics{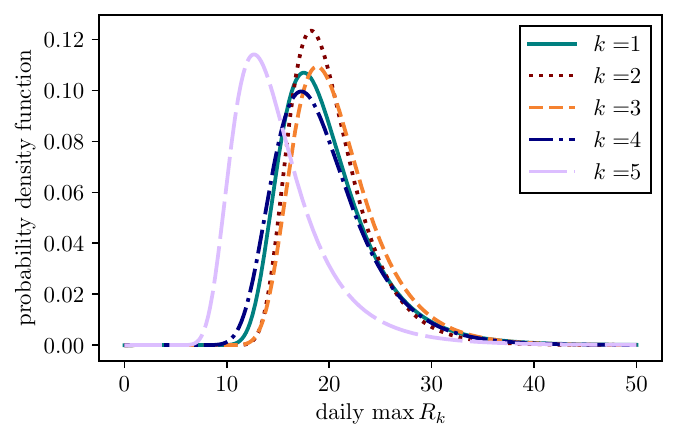}
    \caption{Probability densities $g_{\gamma,b,a}^{\text{EV}}$ for maxima of the first 5 modes, $\hat\gamma^+, \hat\sigma^+$ estimated at a threshold of $u=15$ with POT.}
    \label{fig:pos_extreme_gev}
\end{figure}

\begin{figure}
    \centering  
    \includegraphics{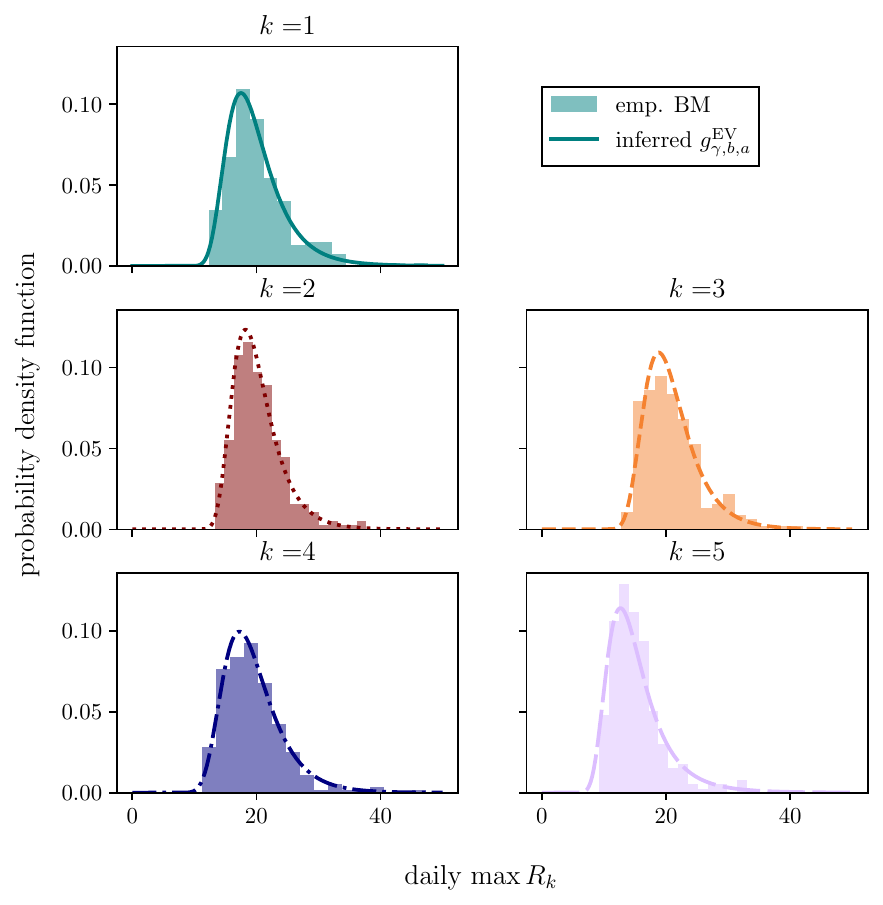}
    \caption{Probability densities $g_{\gamma,b,a}^{\text{EV}}$ for maxima of the first 5 modes, $\hat\gamma^+, \hat\sigma^+$ estimated at a threshold of $u=15$ with POT in comparison with the empirical probability densities obtained via block maxima.}
    \label{fig:pos_extreme_gev_inferred}
\end{figure}

We stress that the strength of our approach lies in reliably estimating $\gamma$, but in addition it is even possible to calculate the parameters $a$ and $b$ for the corresponding probability density \eqref{eq:GEV_density} using \eqref{eq:linearparameterrelation} and by matching the return periods as described in Refs.~\cite{Coles2001, DingGPDGEVRelationship2008}. We note that the probability densities are then theoretically inferred densities of block maxima, \emph{not} measured block maxima densities. We show the probability densities for the daily maximum in Fig.~\ref{fig:pos_extreme_gev}, estimated via the GPD fit at an exemplary threshold of $u=15$. To prove that the calculation of the parameters $a$ and $b$ gives indeed the correct result, we show the empirical daily block maxima densities together with our inferred densities in Fig.~\ref{fig:pos_extreme_gev_inferred}. The distributions fit remarkably well. This is to the best of our knowledge the first empirical analysis of this depth, showing the theoretical correspondence of the POT approach to block maxima methods empirically. It proves that we can avoid the block maxima method entirely, since all information about maxima and minima are contained in the tail of the density. We note that until now, our methods do not treat the nonstationarity of the time series directly. Thus the resulting extreme value statistics should be understood as results of averaging over these effects.

Up to this point we developed the general framework of our analysis and applied it to the rotated modes of financial data. We applied the threshold-based approach to estimate tail behavior and we proved that it allows us to infer also the statistics of the maximum. This checks already three of the four guiding points layed out in the introduction of this paper. We have also begun to address the remaining point --- nonstationarity --- by demonstrating that seasonalities and extremal clustering are present in our data.

\FloatBarrier
\section{Residual Risk and Local Threshold Estimation \label{s:local_estimation}}
This section deals with an estimation for the same modes as in Sec.~\ref{s:dataanalysisresults}, but accounting for nonstationarity and seasonality in the form of volatility clustering, referring to the fourth and last point of the paper's guiding questions. Section~ \ref{ss:local_estimation_results} discusses the results while Sec.~\ref{ss:local_estimation_interpretation} sheds light on the difference to Sec.~\ref{s:dataanalysisresults} in interpretation.
\subsection{Results\label{ss:local_estimation_results}}
As shown in Sec.~\ref{s:dataanalysisresults}, there exist strong intraday seasonality effects, in addition to the usual persistence of the volatility autocorrelations. To distinguish between the more deterministic, seasonal risk and the really unpredictable, residual risk, we calculate the mean intraday volatility profile of the modes as \cite{LiuVolatility1999, PlerouFluctuations2005, WangScalingMemoryIntraday2006}
\begin{align}
    V_k(t_{\text{intra}}) = \frac{1}{N_\text{days}}\sum_{d=0}^{N_\text{days}-1} \vert R_k(d\,T_\text{day}+t_{\text{intra}}) \vert\,, && t_\text{intra} = 1, \ldots, T_\text{day}\, .
\end{align}
Here, $t_\text{intra}$ denotes the intraday time index in seconds. We compute the profile by aligning all trading days and averaging the absolute returns at the same intraday time.

\begin{figure}
    \centering
    \begin{subfigure}{\textwidth}
            \includegraphics[width=\textwidth]{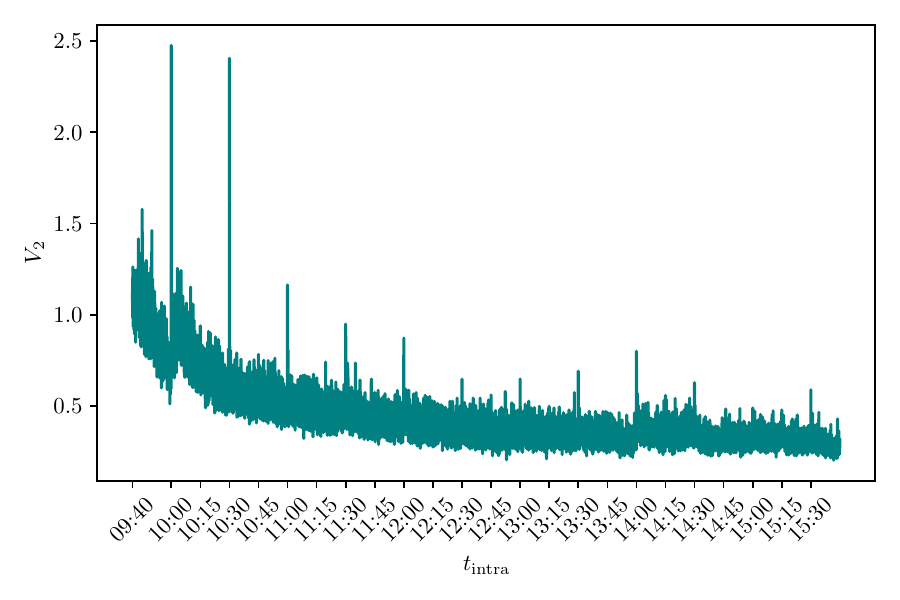}
    \end{subfigure}%
    \\
    \begin{subfigure}{\textwidth}
            \includegraphics[width=\textwidth]{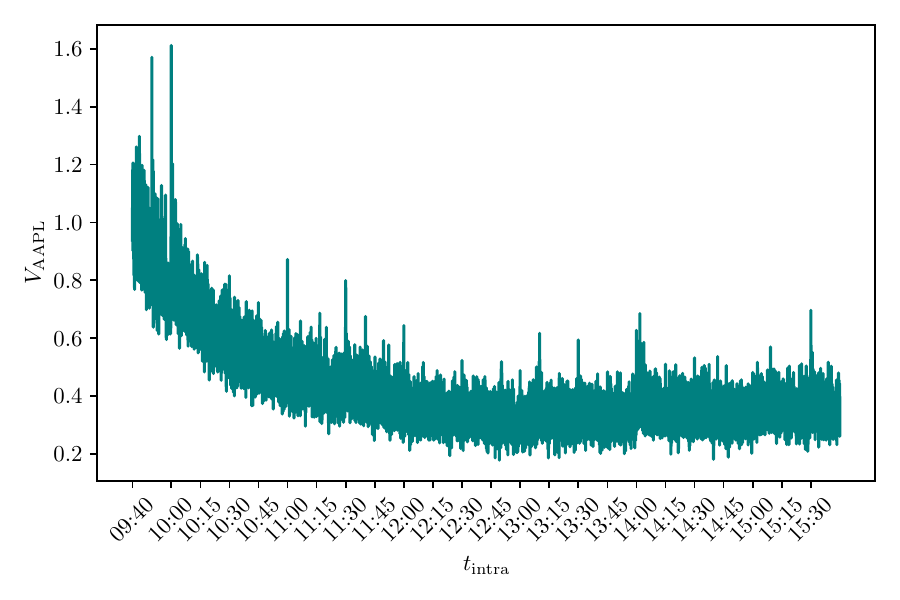}
    \end{subfigure}
    \caption{Top: Volatility profile of the second mode $V_2$ referring to the Energy sector. Bottom: Volatility profile of the stock returns of AAPL $V_\text{AAPL}$.}
    \label{fig:daily_profiles}
\end{figure}

\begin{figure}
    \centering
    \includegraphics{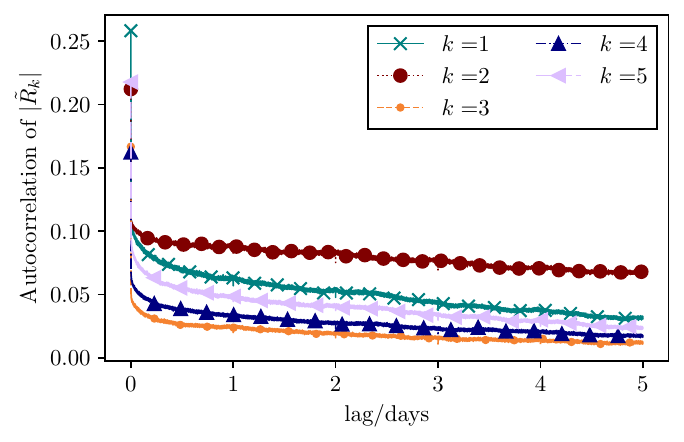}
    \caption{Autocorrelation of absolute values residual time series of rotated return modes $\vert \tilde R_k\vert $, against time lag. The first value for each mode is plotted at a lag of 1s. Markers are evenly spaced for better readability, data frequency is 1s.}
    \label{fig:removed_profile}
\end{figure}

\begin{figure}
    \centering
    \includegraphics[width=0.75\textwidth]{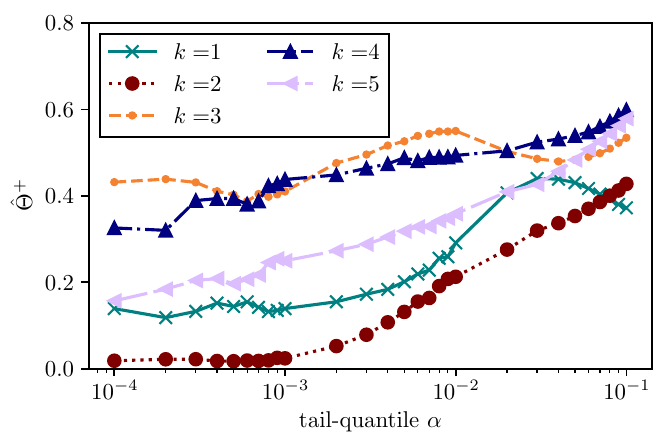}
    \caption{Estimated extremal index $\hat \Theta^+$ of $\tilde R_k$ against tail-quantile $\alpha$.}
    \label{fig:removed_profile_theta_hat_pos}
\end{figure}

Volatility in the beginning and end of a trading day is generally higher, see for instance Refs.~\cite{HarrisIntraday1986, AdmatiIntradayPatterns1988, LiuVolatility1999, WangScalingMemoryIntraday2006}. In addition, at a resolution of seconds, spikes occur at distinct times of the day, see Fig.~\ref{fig:daily_profiles}.
As discussed in Sec.~\ref{ss:data_processing}, spikes would occur also in the end of the trading day, starting at 15:45:00 when information regarding closing auctions is shared with all traders. However, there are similarly sharp spikes throughout the day visible in the volatility profiles for all modes and even for single stocks. Most of these spikes happen at half hour changes, and they are rather stationary throughout the whole year. Due to their sharpness they quickly wash out and are thus not visible in most documented intraday volatility profiles with return horizons of minutes \cite{HarrisIntraday1986, LiuVolatility1999, WangScalingMemoryIntraday2006}. To the best of our knowledge, there is no documented recurring official information release at these times throughout the day. Therefore it is probable that these seconds of increased volatility emerge due to certain trading or rebalancing algorithms and are thus part of natural trading dynamics. Since these spikes are rather predictable in their occurrence and naturally lead to extreme values in returns, it is of interest to distinguish the natural, residual risk of trading from the systematic, regularly occurring risk, especially in our setting of high frequency. We calculate the residual time series of the modes by dividing the returns of the mode by the intraday volatility profile \cite{LiuVolatility1999, WangScalingMemoryIntraday2006}
\begin{align}
    \tilde R_k(t) = R_k(t)/V_k(t \bmod{T_\text{day}})\,, && t=1, \ldots, T.
\end{align}
When evaluating the extremal index and the autocorrelation of $\tilde R_k(t)$ we see that the intraday seasonality is removed, and that it is not responsible for the serial clustering of extremes, see Figs.~\ref{fig:removed_profile} and \ref{fig:removed_profile_theta_hat_pos}.%
On the contrary, significant autocorrelation is still apparent and extremes are clustering very strongly.

\begin{figure}[ht]
    \centering
    \includegraphics[width=\textwidth]{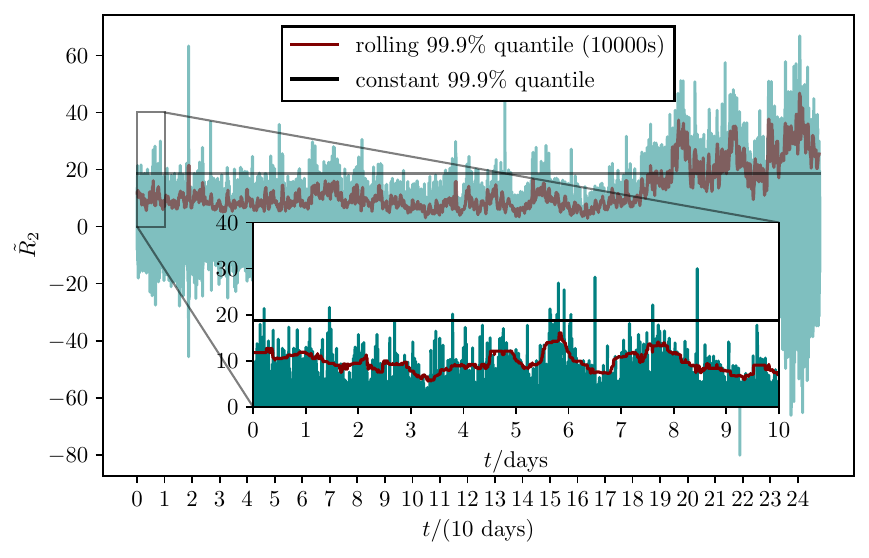}
    \caption{Residuals of mode 2, constant 99.9\% quantile and the local rolling 99.9\% quantile calculated over the last 10000s.}
    \label{fig:clustering_time}
\end{figure}

Thus rather than quantifying the shape parameter of the distribution with a fixed threshold $u$, we apply a dynamic threshold $u(t)$ evaluated by estimating local quantiles with a rolling window. This takes into account the nonstationarity of the return time series, even exogenous effects are incorporated into the rolling window after a few seconds. We have to ensure the window is large enough to contain data points which exceed the local threshold implicitly set by the quantile, since otherwise we will not have any exceedances overall. On the other hand we want to cover the effect of nonstationarity, which limits the length of the window. However, the choice of window size is robust in a range of different lengths, as long as $\alpha$ multiplied with the window size is at the scale of one or larger. In that case we have on average one exceedance per window and can reliably estimate the GPD parameters. We find that a window of 10000s, roughly half a trading day, provides a good balance between these aspects. This length allows us to measure extremes up to a quantile of $99.99\%$ while still accounting for nonstationarity. Displaying the local 99.9\%-quantile over a rolling window of 10000s in Fig.~\ref{fig:clustering_time}, this nonstationarity is very apparent. Using constant thresholds leads inevitably to clustering of extremes in such cases, as the definition of what is ``extreme'' changes with the change of the current state of the market.

\begin{figure}[htb]
    \centering
    \begin{subfigure}{\textwidth}
    	\centering
        \includegraphics[width=0.75\textwidth]{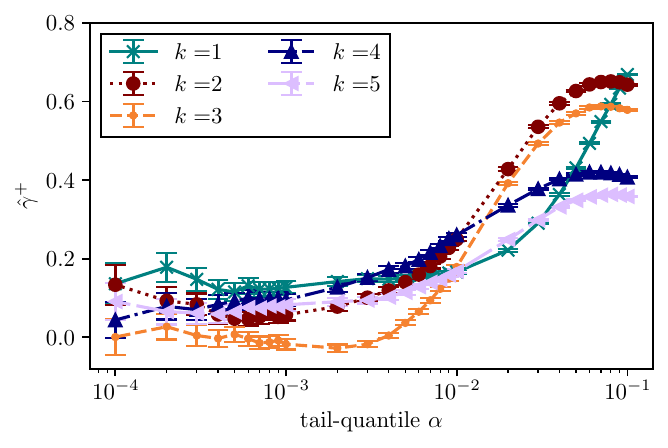}
    \end{subfigure}\\
    \begin{subfigure}{0.5\textwidth}
 \includegraphics[width=\textwidth]{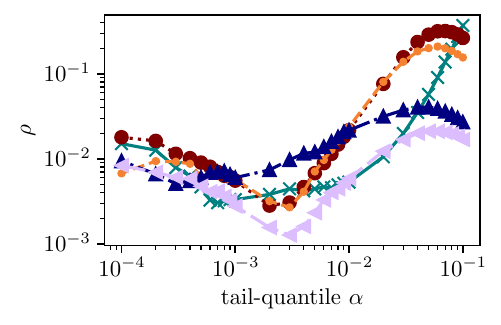}
    \end{subfigure}%
    \begin{subfigure}{0.5\textwidth}
        \includegraphics[width=\textwidth]{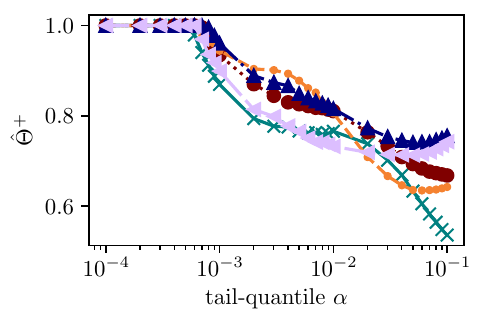}
    \end{subfigure}
    \caption{Estimation of positive tail shape parameter $\hat\gamma^+$ with a rolling threshold determined by the local tail-quantile $\alpha$ over 10000s. Top: Estimated $\hat \gamma$, standard errors according to \eqref{eq:asymptoticvariances}. Bottom left: NRMSD of estimation. Bottom right: Estimated extremal index $\hat \Theta^+$.}
    \label{fig:rolling_estimation+}
\end{figure}

In Fig.~\ref{fig:rolling_estimation+}, the clustering index is plotted against the tail-quantile $\alpha$, which is estimated in a rolling window. Such a local threshold has to remove any serial dependence and autocorrelations by design, and we notice that $\hat \Theta^+$ is considerably higher than for the estimation with nonlocal thresholds. As the threshold tends to tail-quantiles which are high compared to the window size of 10000 data points, the serial dependence is vanishing completely. This is clearly a feature of the rolling window approach, since extremes which lie outside the current window, are not related to extremes inside the current window. In contrast, the removal of the intraday seasonality profile could not alleviate the clustering in any way, as discussed before and shown in Figs.~\ref{fig:removed_profile} and \ref{fig:removed_profile_theta_hat_pos}. In other words the clustering of extremes is not caused by a fixed repeating intraday pattern.

We also notice from the NRMSD in Fig.~\ref{fig:rolling_estimation+} that the range of the goodness-of-fit is roughly equal to the range of the goodness-of-fit for the estimation method using fixed $u$.
The distributions obtained with our local threshold method applied to the residuals of the modes converge to a generalized Pareto distribution, justifying our approach mathematically. Our results using a local threshold and the deseasonalized residuals of the modes differ quantitatively from the results of the method using fixed $u$. Although the range of the tail index is roughly the same, we see that the ordering of the modes indeed changes. Not all modes are influenced with the same strength by the effects of nonstationarity and intraday volatility clustering, thus leading to these changes in order.

\begin{figure}
    \centering
    \includegraphics{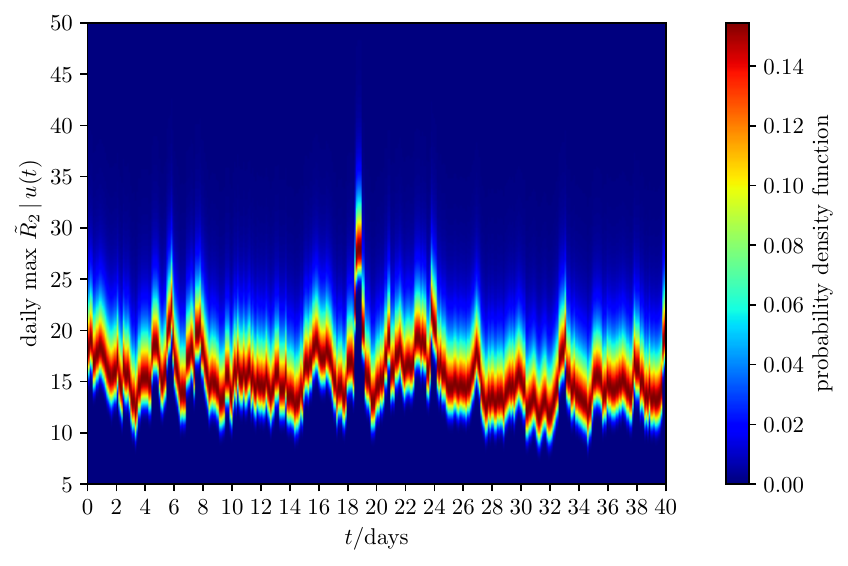}
    \caption{Nonstationary probability density $g_{\gamma,b(t),a}^{\text{EV}}$ of the second mode plotted for forty consecutive trading days. Estimated via local rolling 99.9\% quantile calculated over the last 10000s. Time resolution is 1s, the probability density function is normalized in each time step $t$.}
    \label{fig:pos_extreme_gev_nonstationary}
\end{figure}

As in Sec.~\ref{s:dataanalysisresults}, it is possible to calculate the parameters $a$ and $b$ of \eqref{eq:GEV_density}. Since we use a dynamic threshold $u(t)$ here our location parameter $b(t)$ is time dependent. We show the nonstationary probability density $g_{\gamma,b(t),a}^{\text{EV}}$ for the positive extremes of the second mode as an example in a time range of forty days in Fig.~\ref{fig:pos_extreme_gev_nonstationary}. What we see is that the density of observing a daily extreme, given the current local threshold $u(t)$, changes location drastically in time.

\subsection{Interpretation \label{ss:local_estimation_interpretation}}
When interpreting the results obtained in Secs.~\ref{s:dataanalysisresults} and~\ref{s:local_estimation}, it is crucial to distinguish the different scenarios. In both scenarios we analyzed the return modes, \textit{i.e.}, the return time series of the linear combinations referring to the market and the sectors. These capture the most important dynamics in the large multivariate financial system.

In Sec.~\ref{s:dataanalysisresults}, we analyzed the purely marginal distributions of the modes in one year. This allows us to make statements about the risk profile of the market and its sectors on an annual basis. The quantified risk here includes the rather deterministic surges of volatility and the seasonality effects we showed in Sec.~\ref{s:local_estimation} and does not take into account a nonstationary notion of risk aversion. The threshold of what counts as ``extreme'' in this estimation is set in hindsight and remains fixed, even when the markets themselves become more volatile.

In Sec.~\ref{s:local_estimation}, the perspective is different. Here, we quantify the underlying residual risk while explicitly accounting for the known seasonality profile and the changing risk profile in the state of the market itself. When trading at high frequencies, risk can only be evaluated according to the current market state. For trading strategies that aim at small profits on timescales of seconds, it is not sensible to evaluate risks relative to an annual benchmark. Risk factors of seasonality and known recurring volatility bursts are accounted for beforehand. A sharp peak in volatility for instance is not unusal but expected in the morning of a trading day, and is thus much less an ``extreme'' event than the same peak during midday. So we take the actual residual risk and evaluate it in comparison to the benchmark of the market risk, which is measured locally through a rolling quantile estimation. We see that local thresholds make the notion of an ``extreme'' adaptive to the current market state and substantially reduce spurious clustering induced by regime shifts. This addresses the fourth --- and final --- guiding point of this paper.

We proved that our general approach works well and that it can be adapted to different scenarios.


\FloatBarrier

\section{Conclusion \label{s:conclusion}}
Quantifying risk in complex systems --- exemplary in financial markets where dependencies exist across time, across system constituents, and under nonstationary conditions --- is crucial for understanding and anticipating systemic vulnerabilities. It is also essential for the evaluation of extreme event likelihoods and the design of robust strategies for risk management.

We addressed four guiding points laid out in the introduction. \emph{First}, we developed a general framework for extreme-value analysis in multivariate, strongly correlated time series by reducing the problem to a set of interpretable collective coordinates. \emph{Second}, we rotated high-frequency stock returns into the eigenbasis of the correlation matrix, yielding modes that can be interpreted as collective dynamics of the market and sectoral activity. \emph{Third}, we used the POT approach based on a GPD fit to uncover tail behaviour and, via the correspondence between the GPD and the extreme value distribution, to infer the implied statistics of block maxima without relying on a block maxima estimation procedure. In contrast to the block maxima method, the POT approach is more efficient in data usage and it does not rely on a fixed compartmentalization in the time domain. \emph{Fourth}, we explicitly accounted for nonstationarity by removing intraday seasonality through volatility profiling and by employing local, time dependent thresholds estimated from rolling quantiles to obtain a market state dependent notion of extremes.

To the best of our knowledge, this is the first systematic application of extreme value theory to return modes with frequencies on the scale of seconds obtained from an eigenbasis decomposition, which allows disentangling collective and sectoral sources of risk.

Our results for financial data extend and are consistent with other analyses of tail behavior, usually done for higher return horizons and for indices and single stocks. We show that the extreme value distributions of the market and the sectors lie qualitatively in the Fréchet domain, at high frequency horizons of seconds. Quantitatively, the tail shape parameters differ. Strengths of serial dependencies also differ strongly between modes, with the market and the Energy sector displaying the highest amount of clustering for extreme events. We analyzed the residuals of the return modes, by weighting the time series according to their average intraday volatility profile. This allowed us to focus on the pure stochastic risk. Our analysis proved that the serial dependence is not due to the intraday seasonalities and intraday volatility clustering, but rather due to dependencies on longer time scales and nonstationarities in the volatility. Using local thresholds, we estimate extremal behavior conditional on the current market state. From this local perspective, the tails remain of Fréchet type, but the relative susceptibility of modes to extremes changes due to differing levels of dependence and nonstationarity.

Since our approach is general and not limited to linear measures such as the Pearson correlation used in our example study, the analysis can be extended to include measures focusing directly on extreme values such as matrices of pairwise tail dependencies \textit{et cetera}. Nonlinear measures may be able to order stocks according to their similarity regarding only their extremal behavior. Importantly the method is not limited to financial data and there is a plethora of multivariate systems where such an analysis can provide value in understanding the dynamics and statistics of collective extremal behavior.

\section*{Acknowledgements}
We thank Holger Kantz for fruitful discussions.

\newpage
\printbibliography

@article{OverviewExtremalIndex2019,
  title = {{An Overview of the Extremal Index}},
  author = {Moloney, Nicholas R. and Faranda, Davide and Sato, Yuzuru},
  date = {2019-02-11},
  journaltitle = {Chaos: An Interdisciplinary Journal of Nonlinear Science},
  shortjournal = {Chaos: An Interdisciplinary Journal of Nonlinear Science},
  volume = {29},
  number = {2},
  pages = {022101},
  issn = {1054-1500},
  doi = {10.1063/1.5079656},
  url = {https://doi.org/10.1063/1.5079656},
  urldate = {2025-04-14}
}

@article{FerroSegers2003,
  title = {{Inference for {{Clusters}} of {{Extreme Values}}}},
  author = {Ferro, Christopher A. T. and Segers, Johan},
  year = {2003},
  month = may,
  journal = {Journal of the Royal Statistical Society Series B: Statistical Methodology},
  volume = {65},
  number = {2},
  pages = {545--556},
  issn = {1369-7412, 1467-9868},
  doi = {10.1111/1467-9868.00401},
  urldate = {2025-02-17},
  copyright = {https://academic.oup.com/journals/pages/open\_access/funder\_policies/chorus/standard\_publication\_model},
  langid = {english}
}

@article{Hosking1987,
  title = {Parameter and {{Quantile Estimation}} for the {{Generalized Pareto Distribution}}},
  author = {Hosking, J. R. M. and Wallis, J. R.},
  year = {1987},
  month = aug,
  journal = {Technometrics},
  volume = {29},
  number = {3},
  pages = {339--349},
  publisher = {ASA Website},
  issn = {0040-1706},
  doi = {10.1080/00401706.1987.10488243},
  urldate = {2025-04-14}
}

@incollection{Smith1984,
  title = {Threshold {{Methods}} for {{Sample Extremes}}},
  booktitle = {Statistical {{Extremes}} and {{Applications}}},
  author = {Smith, Richard L.},
  editor = {{de Oliveira}, J. Tiago},
  year = {1984},
  pages = {621--638},
  publisher = {Springer Netherlands},
  address = {Dordrecht},
  doi = {10.1007/978-94-017-3069-3_48},
  urldate = {2025-04-14},
  isbn = {978-94-017-3069-3},
  langid = {english}
}

@book{haan2006,
  title={{Extreme Value Theory: An Introduction}},
  author={Haan, Laurens and Ferreira, Ana},
  volume={3},
  year={2006},
  publisher={Springer}
}

@book{Coles2001,
  title = {An {{Introduction}} to {{Statistical Modeling}} of {{Extreme Values}}},
  author = {Coles, Stuart},
  year = {2001},
  series = {Springer {{Series}} in {{Statistics}}},
  publisher = {Springer London},
  address = {London},
  doi = {10.1007/978-1-4471-3675-0},
  urldate = {2025-04-14},
  copyright = {http://www.springer.com/tdm},
  isbn = {978-1-84996-874-4 978-1-4471-3675-0},
  langid = {english}
}

@book{Majumdar2024,
  title = {Statistics of {{Extremes}} and {{Records}} in {{Random Sequences}}},
  author = {Majumdar, Satya N. and Schehr, Gr{\'e}gory},
  year = {2024},
  month = jul,
  publisher = {Oxford University Press},
  googlebooks = {JyMJEQAAQBAJ},
  isbn = {978-0-19-251788-3},
  langid = {english}
}

@article{Leadbetter1974,
  title = {{On Extreme Values in Stationary Sequences}},
  author = {Leadbetter, M. R.},
  year = {1974},
  journal = {Zeitschrift für Wahr\-schein\-lich\-keits\-theorie und Verwandte Gebiete},
  volume = {28},
  number = {4},
  pages = {289--303},
  issn = {0044-3719, 1432-2064},
  doi = {10.1007/BF00532947},
  urldate = {2025-03-28},
  copyright = {http://www.springer.com/tdm},
  langid = {english}
}

@article{Leadbetter1983,
  title = {{Extremes and Local Dependence in Stationary Sequences}},
  author = {Leadbetter, M R},
  journal={Zeitschrift f{\"u}r Wahrscheinlichkeitstheorie und verwandte Gebiete},
  volume={65},
  pages={291--306},
  year={1983},
  publisher={Springer-Verlag},
  langid = {english}
}

@article{GeneralPerspectiveWeather2011,
  title = {{A General Perspective of Extreme Events in Weather and Climate}},
  author = {Sura, Philip},
  year = {2011},
  month = jul,
  journal = {Atmospheric Research},
  volume = {101},
  number = {1},
  pages = {1--21},
  issn = {0169-8095},
  doi = {10.1016/j.atmosres.2011.01.012},
  urldate = {2025-05-05}
}

@article{ClimateChange2014,
  title = {{Non-Stationary Extreme Value Analysis in a Changing Climate}},
  author = {Cheng, Linyin and AghaKouchak, Amir and Gilleland, Eric and Katz, Richard W.},
  year = {2014},
  month = nov,
  journal = {Climatic Change},
  volume = {127},
  number = {2},
  pages = {353--369},
  issn = {1573-1480},
  doi = {10.1007/s10584-014-1254-5},
  urldate = {2025-05-05}
}

@article{HydrologyPaper2004,
  title = {Practical {{Extreme Value Modelling}} of {{Hydrological Floods}} and {{Droughts}}: {{A Case Study}}},
  shorttitle = {Practical {{Extreme Value Modelling}} of {{Hydrological Floods}} and {{Droughts}}},
  author = {Engeland, Kolbj{\o}rn and Hisdal, Hege and Frigessi, Arnoldo},
  year = {2004},
  month = mar,
  journal = {Extremes},
  volume = {7},
  number = {1},
  pages = {5--30},
  issn = {1572-915X},
  doi = {10.1007/s10687-004-4727-5},
  urldate = {2025-05-05},
}

@article{SafetyEstimation2006,
  title = {{The Extreme Value Theory Approach to Safety Estimation}},
  author = {Songchitruksa, Praprut and Tarko, Andrew P.},
  year = {2006},
  month = jul,
  journal = {Accident Analysis \& Prevention},
  volume = {38},
  number = {4},
  pages = {811--822},
  issn = {0001-4575},
  doi = {10.1016/j.aap.2006.02.003},
  urldate = {2025-05-05}
}

@article{Münnix2012,
  title = {Identifying {{States}} of a {{Financial Market}}},
  author = {M{\"u}nnix, Michael C. and Shimada, Takashi and Sch{\"a}fer, Rudi and Leyvraz, Francois and Seligman, Thomas H. and Guhr, Thomas and Stanley, H. Eugene},
  year = {2012},
  month = sep,
  journal = {Scientific Reports},
  volume = {2},
  number = {1},
  pages = {644},
  publisher = {Nature Publishing Group},
  issn = {2045-2322},
  doi = {10.1038/srep00644},
  urldate = {2025-05-05},
}

@article{FisherTippett1928,
  title = {{Limiting Forms of the Frequency Distribution of the Largest or Smallest Member of a Sample}},
  author = {Fisher, R. A. and Tippett, L. H. C.},
  year = {1928},
  month = apr,
  journal = {Mathematical Proceedings of the Cambridge Philosophical Society},
  volume = {24},
  number = {2},
  pages = {180--190},
  issn = {1469-8064, 0305-0041},
  doi = {10.1017/S0305004100015681},
  urldate = {2025-05-05},
}

@article{Gnedenko1943,
  title = {Sur {{La Distribution Limite Du Terme Maximum D}}'{{Une S{\'e}rie Al{\'e}atoire}}},
  author = {Gnedenko, B.},
  year = {1943},
  journal = {Annals of Mathematics},
  volume = {44},
  number = {3},
  eprint = {1968974},
  eprinttype = {jstor},
  pages = {423--453},
  publisher = {[Annals of Mathematics, Trustees of Princeton University on Behalf of the Annals of Mathematics, Mathematics Department, Princeton University]},
  issn = {0003-486X},
  doi = {10.2307/1968974},
  urldate = {2025-05-05}
}

@article{Pickands1975,
  title = {Statistical {{Inference Using Extreme Order Statistics}}},
  author = {Pickands III, James},
  year = {1975},
  month = jan,
  journal = {The Annals of Statistics},
  volume = {3},
  number = {1},
  pages = {119--131},
  publisher = {Institute of Mathematical Statistics},
  issn = {0090-5364, 2168-8966},
  doi = {10.1214/aos/1176343003},
  urldate = {2025-05-05}
}

@article{BalkemaDehaan1974,
  title = {Residual {{Life Time}} at {{Great Age}}},
  author = {Balkema, A. A. and de Haan, L.},
  year = {1974},
  month = oct,
  journal = {The Annals of Probability},
  volume = {2},
  number = {5},
  pages = {792--804},
  publisher = {Institute of Mathematical Statistics},
  issn = {0091-1798, 2168-894X},
  doi = {10.1214/aop/1176996548},
  urldate = {2025-05-05},
}

@article{DiscussionDavison1990,
  title = {Discussion of the {{Paper}} by {{Davison}} and {{Smith}}},
  year = {1990},
  month = jan,
  journal = {Journal of the Royal Statistical Society Series B: Statistical Methodology},
  volume = {52},
  number = {3},
  pages = {425--442},
  issn = {0035-9246},
  doi = {10.1111/j.2517-6161.1990.tb01797.x},
  urldate = {2025-03-25}
}

@article{Salmon2012,
  title = {The {{Formula}} That {{Killed Wall Street}}},
  author = {Salmon, Felix},
  year = {2012},
  month = feb,
  journal = {Significance},
  volume = {9},
  number = {1},
  pages = {16--20},
  issn = {1740-9705},
  doi = {10.1111/j.1740-9713.2012.00538.x},
  urldate = {2025-05-05}
}

@article{Ding1993,
  title = {{A Long Memory Property of Stock Market Returns and a New Model}},
  author = {Ding, Zhuanxin and Granger, Clive W. J. and Engle, Robert F.},
  year = {1993},
  month = jun,
  journal = {Journal of Empirical Finance},
  volume = {1},
  number = {1},
  pages = {83--106},
  issn = {0927-5398},
  doi = {10.1016/0927-5398(93)90006-D},
  urldate = {2025-05-06}
}

@article{Lux1996,
  title = {{Long-Term Stochastic Dependence in Financial Prices: Evidence from the {{German}} Stock Market}},
  shorttitle = {Long-Term Stochastic Dependence in Financial Prices},
  author = {Lux, Thomas},
  year = {1996},
  month = nov,
  journal = {Applied Economics Letters},
  volume = {3},
  number = {11},
  pages = {701--706},
  publisher = {Routledge},
  issn = {1350-4851},
  doi = {10.1080/135048596355691},
  urldate = {2025-05-06}
}

@article{Markowitz1952,
  title = {Portfolio {{Selection}}},
  author = {Markowitz, Harry},
  year = {1952},
  journal = {The Journal of Finance},
  volume = {7},
  number = {1},
  eprint = {2975974},
  eprinttype = {jstor},
  pages = {77--91},
  publisher = {[American Finance Association, Wiley]},
  issn = {0022-1082},
  doi = {10.2307/2975974},
  urldate = {2025-05-06}
}

@article{Laloux1999,
  title = {Noise {{Dressing}} of {{Financial Correlation Matrices}}},
  author = {Laloux, Laurent and Cizeau, Pierre and Bouchaud, Jean-Philippe and Potters, Marc},
  year = {1999},
  month = aug,
  journal = {Physical Review Letters},
  volume = {83},
  number = {7},
  pages = {1467--1470},
  issn = {0031-9007, 1079-7114},
  doi = {10.1103/PhysRevLett.83.1467},
  urldate = {2025-05-06},
  copyright = {http://link.aps.org/licenses/aps-default-license},
}

@article{Davison1990,
  title = {Models for {{Exceedances Over High Thresholds}}},
  author = {Davison, A. C. and Smith, R. L.},
  year = {1990},
  month = jul,
  journal = {Journal of the Royal Statistical Society Series B: Statistical Methodology},
  volume = {52},
  number = {3},
  pages = {393--425},
  issn = {1369-7412, 1467-9868},
  doi = {10.1111/j.2517-6161.1990.tb01796.x},
  urldate = {2025-03-25},
  copyright = {https://academic.oup.com/journals/pages/open\_access/funder\_policies/chorus/standard\_publication\_model}
}

@article{Plerou2000,
  title = {{A Random Matrix Theory Approach to Financial Cross-Correlations}},
  author = {Plerou, V and Gopikrishnan, P and Rosenow, B and Amaral, L. A. N and Stanley, H. E},
  year = {2000},
  month = dec,
  journal = {Physica A: Statistical Mechanics and its Applications},
  volume = {287},
  number = {3},
  pages = {374--382},
  issn = {0378-4371},
  doi = {10.1016/S0378-4371(00)00376-9},
  urldate = {2025-05-13}
}

@article{Koedijk1990,
  title = {{The Tail Index of Exchange Rate Returns}},
  author = {Koedijk, Kees G. and Schafgans, Marcia M. A. and {de Vries}, Casper G.},
  year = {1990},
  month = aug,
  journal = {Journal of International Economics},
  volume = {29},
  number = {1},
  pages = {93--108},
  issn = {0022-1996},
  doi = {10.1016/0022-1996(90)90065-T},
  urldate = {2025-05-13}
}

@article{Münnix2011,
  title = {{A Copula Approach on the Dynamics of Statistical Dependencies in the {{US}} Stock Market}},
  author = {Münnix, Michael C. and Schäfer, Rudi},
  year = {2011},
  month = nov,
  journal = {Physica A: Statistical Mechanics and its Applications},
  volume = {390},
  number = {23},
  pages = {4251--4259},
  issn = {0378-4371},
  doi = {10.1016/j.physa.2011.06.032},
  urldate = {2025-05-14}
}

@article{Epps1979,
	title = {Comovements in {{Stock Prices}} in the {{Very Short Run}}},
	author = {Epps, Thomas W.},
	year = {1979},
	journal = {Journal of the American Statistical Association},
	volume = {74},
	number = {366},
	eprint = {2286325},
	eprinttype = {jstor},
	pages = {291--298},
	publisher = {[American Statistical Association, Taylor \& Francis, Ltd.]},
	issn = {0162-1459},
	doi = {10.2307/2286325},
	urldate = {2025-07-03}
}

@article{HayashiYoshida2005,
	title = {On Covariance Estimation of Non-Synchronously Observed Diffusion Processes},
	author = {Hayashi, Takaki and Yoshida, Nakahiro},
	year = {2005},
	month = apr,
	journal = {Bernoulli},
	volume = {11},
	number = {2},
	pages = {359--379},
	publisher = {{Bernoulli Society for Mathematical Statistics and Probability}},
	issn = {1350-7265},
	doi = {10.3150/bj/1116340299},
	urldate = {2025-07-03},
}

@article{Münnix2010,
	title = {Compensating Asynchrony Effects in the Calculation of Financial Correlations},
	author = {M{\"u}nnix, Michael C. and Sch{\"a}fer, Rudi and Guhr, Thomas},
	year = {2010},
	month = feb,
	journal = {Physica A: Statistical Mechanics and its Applications},
	volume = {389},
	number = {4},
	pages = {767--779},
	issn = {0378-4371},
	doi = {10.1016/j.physa.2009.10.033},
	urldate = {2025-07-03},
}

@article{MünnixTickSize2010,
	title = {Impact of the Tick-Size on Financial Returns and Correlations},
	author = {M{\"u}nnix, Michael C. and Sch{\"a}fer, Rudi and Guhr, Thomas},
	year = {2010},
	month = nov,
	journal = {Physica A: Statistical Mechanics and its Applications},
	volume = {389},
	number = {21},
	pages = {4828--4843},
	issn = {0378-4371},
	doi = {10.1016/j.physa.2010.06.037},
	urldate = {2025-07-03},
}

@book{Weibull1939,
	title = {A {{Statistical Theory}} of the {{Strength}} of {{Materials}}},
	author = {Weibull, Waloddi},
	year = {1939},
	publisher = {Generalstabens li\-to\-gra\-fis\-ka anstalts f{\"o}rlag},
	googlebooks = {otVRAQAAIAAJ},
	langid = {english}
}

@article{GumbelReturnPeriod1941,
	title = {The {{Return Period}} of {{Flood Flows}}},
	author = {Gumbel, E. J.},
	year = {1941},
	journal = {The Annals of Mathematical Statistics},
	volume = {12},
	number = {2},
	eprint = {2235766},
	eprinttype = {jstor},
	pages = {163--190},
	publisher = {Institute of Mathematical Statistics},
	issn = {0003-4851},
	urldate = {2025-07-03},
}

@article{Gumbel1942,
	title = {On the {{Frequency Distribution}} of {{Extreme Values}} in {{Meteorological Data}}},
	author = {Gumbel, E. J.},
	year = {1942},
	journal = {Bulletin of the American Meteorological Society},
	volume = {23},
	number = {3},
	eprint = {26256302},
	eprinttype = {jstor},
	pages = {95--105},
	publisher = {American Meteorological Society},
	issn = {0003-0007},
	urldate = {2025-07-03},
}

@article{Plerou1999,
	title = {Scaling of the Distribution of Price Fluctuations of Individual Companies},
	author = {Plerou, Vasiliki and Gopikrishnan, Parameswaran and Nunes Amaral, Lu{\'i}s A. and Meyer, Martin and Stanley, H. Eugene},
	year = {1999},
	month = dec,
	journal = {Physical Review E},
	volume = {60},
	number = {6},
	pages = {6519--6529},
	issn = {1063-651X, 1095-3787},
	doi = {10.1103/PhysRevE.60.6519},
	urldate = {2025-07-03},
	copyright = {http://link.aps.org/licenses/aps-default-license},
	langid = {english},
}

@book{EmbrechtsModellingExtremalEvents1997,
	title = {Modelling {{Extremal Events}}},
	author = {Embrechts, Paul and Kl{\"u}ppelberg, Claudia and Mikosch, Thomas},
	year = {1997},
	publisher = {Springer Berlin Heidelberg},
	address = {Berlin, Heidelberg},
	doi = {10.1007/978-3-642-33483-2},
	urldate = {2025-07-03},
	copyright = {http://www.springer.com/tdm},
	isbn = {978-3-642-08242-9 978-3-642-33483-2},
	langid = {english},
}

@article{Majumdar2020Review,
  title = {Extreme Value Statistics of Correlated Random Variables: {{A}} Pedagogical Review},
  shorttitle = {Extreme Value Statistics of Correlated Random Variables},
  author = {Majumdar, Satya N. and Pal, Arnab and Schehr, Gr{\'e}gory},
  year = {2020},
  month = jan,
  journal = {Physics Reports},
  series = {Extreme Value Statistics of Correlated Random Variables: {{A}} Pedagogical Review},
  volume = {840},
  pages = {1--32},
  issn = {0370-1573},
  doi = {10.1016/j.physrep.2019.10.005},
  urldate = {2025-07-31}
}

@article{GICS,
  title={{Global Industry Classification Sector (GICS)}},
  author={{MSCI, S\&P Dow Jones Indices LLC and its affiliates}},
  year={2023},
  publisher={MSCI}
}

@article{KessyWhitening2018,
  title = {Optimal {{Whitening}} and {{Decorrelation}}},
  author = {Kessy, Agnan and Lewin, Alex and Strimmer, Korbinian},
  year = {2018},
  month = oct,
  journal = {The American Statistician},
  volume = {72},
  number = {4},
  pages = {309--314},
  publisher = {ASA Website},
  issn = {0003-1305},
  doi = {10.1080/00031305.2016.1277159},
  urldate = {2025-07-31}
}

@article{LiuVolatility1999,
  title = {Statistical Properties of the Volatility of Price Fluctuations},
  author = {Liu, Yanhui and Gopikrishnan, Parameswaran and {Cizeau} and {Meyer} and {Peng} and Stanley, H. Eugene},
  date = {1999-08-01},
  journaltitle = {Physical Review E},
  shortjournal = {Phys. Rev. E},
  volume = {60},
  number = {2},
  pages = {1390--1400},
  issn = {1063-651X, 1095-3787},
  doi = {10.1103/PhysRevE.60.1390},
  url = {https://link.aps.org/doi/10.1103/PhysRevE.60.1390},
  urldate = {2025-09-10},
  langid = {english},
}

@article{PlerouFluctuations2005,
  title = {Quantifying Fluctuations in Market Liquidity: {{Analysis}} of the Bid-Ask Spread},
  shorttitle = {Quantifying Fluctuations in Market Liquidity},
  author = {Plerou, Vasiliki and Gopikrishnan, Parameswaran and Stanley, H. Eugene},
  date = {2005-04-22},
  journaltitle = {Physical Review E},
  shortjournal = {Phys. Rev. E},
  volume = {71},
  number = {4},
  pages = {046131},
  issn = {1539-3755, 1550-2376},
  doi = {10.1103/PhysRevE.71.046131},
  url = {https://link.aps.org/doi/10.1103/PhysRevE.71.046131},
  urldate = {2025-09-10},
  langid = {english},
}

@article{AdmatiIntradayPatterns1988,
  title = {A {{Theory}} of {{Intraday Patterns}}: {{Volume}} and {{Price Variability}}},
  shorttitle = {A {{Theory}} of {{Intraday Patterns}}},
  author = {Admati, Anat R. and Pfleiderer, Paul},
  date = {1988-01-01},
  journaltitle = {The Review of Financial Studies},
  shortjournal = {Rev Financ Stud},
  volume = {1},
  number = {1},
  pages = {3--40},
  issn = {0893-9454},
  doi = {10.1093/rfs/1.1.3},
  url = {https://doi.org/10.1093/rfs/1.1.3},
  urldate = {2025-09-10},
}

@article{HarrisIntraday1986,
  title = {A Transaction Data Study of Weekly and Intradaily Patterns in Stock Returns},
  author = {Harris, Lawrence},
  date = {1986-05-01},
  journaltitle = {Journal of Financial Economics},
  shortjournal = {Journal of Financial Economics},
  volume = {16},
  number = {1},
  pages = {99--117},
  issn = {0304-405X},
  doi = {10.1016/0304-405X(86)90044-9},
  url = {https://www.sciencedirect.com/science/article/pii/0304405X86900449},
  urldate = {2025-09-10},

}

@article{WangScalingMemoryIntraday2006,
  title = {Scaling and Memory of Intraday Volatility Return Intervals in Stock Markets},
  author = {Wang, Fengzhong and Yamasaki, Kazuko and Havlin, Shlomo and Stanley, H. Eugene},
  date = {2006-02-16},
  journaltitle = {Physical Review E},
  shortjournal = {Phys. Rev. E},
  volume = {73},
  number = {2},
  pages = {026117},
  issn = {1539-3755, 1550-2376},
  doi = {10.1103/PhysRevE.73.026117},
  url = {https://link.aps.org/doi/10.1103/PhysRevE.73.026117},
  urldate = {2025-09-10},
  langid = {english},
}

@article{BacidoreTradingClose2012,
  title = {Trading around the {{Close}}},
  author = {Bacidore, Jeff and Polidore, Ben and Xu, Wenjie and Yang, Cindy Y.},
  date = {2012-12-31},
  journaltitle = {The Journal of Trading},
  shortjournal = {JOT},
  volume = {8},
  number = {1},
  pages = {48--57},
  issn = {1559-3967, 2168-8427},
  doi = {10.3905/jot.2012.8.1.048},
  url = {http://pm-research.com/lookup/doi/10.3905/jot.2012.8.1.048},
  urldate = {2025-09-10},
}

@online{NYSERegulationRule,
  title = {{{NYSE Regulation}}: {{Rule Interpretations}}},
  url = {https://www.nyse.com/regulation/rule-interpretations},
  urldate = {2025-09-10},
}

@article{GopikrishnanScaling1999,
  title = {Scaling of the Distribution of Fluctuations of Financial Market Indices},
  author = {Gopikrishnan, Parameswaran and Plerou, Vasiliki and Nunes Amaral, Luís A. and Meyer, Martin and Stanley, H. Eugene},
  date = {1999-11-01},
  journaltitle = {Physical Review E},
  shortjournal = {Phys. Rev. E},
  volume = {60},
  number = {5},
  pages = {5305--5316},
  issn = {1063-651X, 1095-3787},
  doi = {10.1103/PhysRevE.60.5305},
  url = {https://link.aps.org/doi/10.1103/PhysRevE.60.5305},
  urldate = {2025-09-12},
  langid = {english},
}

@article{PhillipsKantz2025,
  title = {Improved {{Trend Analysis With EOFs}} and {{Application}} to {{Warming}} of {{Polar Regions}}},
  author = {Phillips, Ewan T. and Kantz, Holger},
  date = {2025},
  journaltitle = {International Journal of Climatology},
  volume = {45},
  number = {7},
  pages = {e8823},
  issn = {1097-0088},
  doi = {10.1002/joc.8823},
  url = {https://onlinelibrary.wiley.com/doi/abs/10.1002/joc.8823},
  urldate = {2025-09-15},
  langid = {english}
}

@article{Livan2012,
  title = {On the Non-Stationarity of Financial Time Series: Impact on Optimal Portfolio Selection},
  shorttitle = {On the Non-Stationarity of Financial Time Series},
  author = {Livan, Giacomo and Inoue, Jun-ichi and Scalas, Enrico},
  date = {2012-07-26},
  journaltitle = {Journal of Statistical Mechanics: Theory and Experiment},
  shortjournal = {J. Stat. Mech.},
  volume = {2012},
  number = {07},
  pages = {P07025},
  issn = {1742-5468},
  doi = {10.1088/1742-5468/2012/07/P07025},
  url = {https://iopscience.iop.org/article/10.1088/1742-5468/2012/07/P07025},
  urldate = {2025-09-15},
  langid = {english},
}

@article{Manolakis_I,
  title = {Multivariate Distributions in Non-Stationary Complex Systems {{I}}: {{A}} Random Matrix Model and Formulae for Data Analysis},
  shorttitle = {Multivariate Distributions in Non-Stationary Complex Systems {{I}}},
  author = {Manolakis, Efstratios and Heckens, Anton J and Köhler, Benjamin and Guhr, Thomas},
  date = {2025-10-01},
  journaltitle = {Journal of Statistical Mechanics: Theory and Experiment},
  shortjournal = {J. Stat. Mech.},
  volume = {2025},
  number = {10},
  pages = {103404},
  issn = {1742-5468},
  doi = {10.1088/1742-5468/ae0317},
  url = {https://iopscience.iop.org/article/10.1088/1742-5468/ae0317},
  urldate = {2025-12-12},
}

@article{Heckens2025_II,
  title = {Multivariate Distributions in Non-Stationary Complex Systems {{II}}: {{Empirical}} Results for Correlated Stock Markets},
  shorttitle = {Multivariate Distributions in Non-Stationary Complex Systems {{II}}},
  author = {Heckens, Anton J and Manolakis, Efstratios and Schuhmann, Cedric and Guhr, Thomas},
  date = {2025-10-01},
  journaltitle = {Journal of Statistical Mechanics: Theory and Experiment},
  shortjournal = {J. Stat. Mech.},
  volume = {2025},
  number = {10},
  pages = {103405},
  issn = {1742-5468},
  doi = {10.1088/1742-5468/ade5f9},
  url = {https://iopscience.iop.org/article/10.1088/1742-5468/ade5f9},
  urldate = {2025-12-12},
}

@article{GuhrSchell2021,
  title = {Exact Multivariate Amplitude Distributions for Non-Stationary {{Gaussian}} or Algebraic Fluctuations of Covariances or Correlations},
  author = {Guhr, Thomas and Schell, Andreas},
  date = {2021-03-26},
  journaltitle = {Journal of Physics A: Mathematical and Theoretical},
  shortjournal = {J. Phys. A: Math. Theor.},
  volume = {54},
  number = {12},
  pages = {125002},
  issn = {1751-8113, 1751-8121},
  doi = {10.1088/1751-8121/abe3c8},
  url = {https://iopscience.iop.org/article/10.1088/1751-8121/abe3c8},
  urldate = {2025-12-12},
}

@article{DingGPDGEVRelationship2008,
  title = {A Newly-Discovered {{GPD-GEV}} Relationship Together with Comparing Their Models of Extreme Precipitation in Summer},
  author = {Ding, Yuguo and Cheng, Bingyan and Jiang, Zhihong},
  date = {2008-05-01},
  journaltitle = {Advances in Atmospheric Sciences},
  shortjournal = {Adv. Atmos. Sci.},
  volume = {25},
  number = {3},
  pages = {507--516},
  issn = {1861-9533},
  doi = {10.1007/s00376-008-0507-5},
  url = {https://doi.org/10.1007/s00376-008-0507-5},
  urldate = {2026-02-23}
}

@article{McNeilFrey2000,
  title = {Estimation of Tail-Related Risk Measures for Heteroscedastic Financial Time Series: An Extreme Value Approach},
  shorttitle = {Estimation of Tail-Related Risk Measures for Heteroscedastic Financial Time Series},
  author = {McNeil, Alexander J. and Frey, Rüdiger},
  date = {2000-11},
  journaltitle = {Journal of Empirical Finance},
  shortjournal = {Journal of Empirical Finance},
  volume = {7},
  number = {3--4},
  pages = {271--300},
  issn = {09275398},
  doi = {10.1016/S0927-5398(00)00012-8},
  url = {https://linkinghub.elsevier.com/retrieve/pii/S0927539800000128},
  urldate = {2026-02-26},
  langid = {english}
}

@article{Longin1996,
  title = {The {{Asymptotic Distribution}} of {{Extreme Stock Market Returns}}},
  author = {Longin, François M.},
  date = {1996},
  journaltitle = {The Journal of Business},
  volume = {69},
  number = {3},
  eprint = {2353373},
  eprinttype = {jstor},
  pages = {383--408},
  publisher = {University of Chicago Press},
  issn = {0021-9398},
  url = {https://www.jstor.org/stable/2353373},
  urldate = {2026-02-26}
}

@article{Micciche2025,
  title = {Role of Correlations in the Maximum Distribution of Strongly Correlated Stationary {{Markovian}} Processes},
  author = {Miccichè, S.},
  date = {2025-03-01},
  journaltitle = {Chaos, Solitons \& Fractals},
  shortjournal = {Chaos, Solitons \& Fractals},
  volume = {192},
  pages = {115995},
  issn = {0960-0779},
  doi = {10.1016/j.chaos.2025.115995},
  url = {https://www.sciencedirect.com/science/article/pii/S0960077925000086},
  urldate = {2026-07-01},
}

\newpage
\appendix
\section[A]{Appendix\label{s:appendix}}
\subsection{List of Tickersymbols (alphabetically sorted)}
A, AA, AAPL, ABBV, ABC, ABT, ACE, ACN, ACT, ADBE, ADI, ADM, ADP, ADS,
ADSK, ADT, AEE, AEP, AES, AET, AFL, AGN, AIG, AIV, AIZ, AKAM, ALL, ALLE,
ALTR, ALXN, AMAT, AME, AMGN, AMP, AMT, AMZN, AN, AON, APA, APC,
APD, APH, ARG, ATI, AVB, AVP, AVY, AXP, AZO, BA, BAC, BAX, BBBY, BBT,
BBY, BCR, BDX, BEN, BHI, BIIB, BK, BLK, BLL, BMY, BRCM, BSX, BWA, BXP,
C, CA, CAG, CAH, CAM, CAT, CB, CBG, CBS, CCE, CCI, CCL, CELG, CERN, CF,
CFN, CHK, CHRW, CI, CINF, CL, CLX, CMA, CMCSA, CME, CMG, CMI, CMS,
CNP, CNX, COF, COG, COH, COL, COP, COST, COV, CPB, CRM, CSC, CSCO,
CSX, CTAS, CTL, CTSH, CTXS, CVC, CVS, CVX, D, DAL, DD, DE, DFS, DG,
DGX, DHI, DHR, DIS, DISCA, DLPH, DLTR, DNB, DNR, DO, DOV, DOW, DPS,
DRI, DTE, DTV, DUK, DVA, DVN, EA, EBAY, ECL, ED, EFX, EIX, EL, EMC,
EMN, EMR, EOG, EQR, EQT, ESRX, ESV, ETFC, ETN, ETR, EW, EXC, EXPD,
EXPE, F, FAST, FB, FCX, FDO, FDX, FE, FFIV, FIS, FISV, FITB, FLIR, FLR,
FLS, FMC, FOSL, FOXA, FSLR, FTI, FTR, GAS, GCI, GD, GE, GGP, GILD, GIS,
GLW, GM, GME, GNW, GOOG, GPC, GPS, GRMN, GS, GT, GWW, HAL, HAR,
HAS, HBAN, HCBK, HCN, HCP, HD, HES, HIG, HOG, HON, HOT, HP, HPQ, HRB,
HRL, HRS, HSP, HST, HSY, HUM, IBM, ICE, IFF, INTC, INTU, IP, IPG, IR, IRM,
ISRG, ITW, IVZ, JCI, JEC, JNJ, JNPR, JOY, JPM, JWN, K, KEY, KIM, KLAC,
KMB, KMI, KMX, KO, KR, KRFT, KSS, KSU, L, LB, LEG, LEN, LH, LIFE, LLL,
LLTC, LLY, LM, LMT, LNC, LO, LOW, LRCX, LUK, LUV, LYB, M, MA, MAC,
MAR, MAS, MAT, MCD, MCHP, MCK, MCO, MDLZ, MDT, MET, MHFI, MHK,
MJN, MKC, MMC, MMM, MNST, MO, MON, MOS, MPC, MRK, MRO, MS, MSFT,
MSI, MTB, MU, MUR, MWV, MYL, NBR, NDAQ, NE, NEE, NEM, NFLX, NFX,
NI, NKE, NLSN, NOC, NOV, NRG, NSC, NTAP, NTRS, NUE, NVDA, NWL, NWSA,
OI, OKE, OMC, ORCL, ORLY, OXY, PAYX, PBCT, PBI, PCAR, PCG, PCL, PCLN,
PCP, PDCO, PEG, PEP, PETM, PFE, PFG, PG, PGR, PH, PHM, PKI, PLD, PLL,
PM, PNC, PNR, PNW, POM, PPG, PPL, PRGO, PRU, PSA, PSX, PVH, PWR, PX,
PXD, QCOM, QEP, R, RAI, RDC, REGN, RF, RHI, RHT, RIG, RL, ROK, ROP,
ROST, RRC, RSG, RTN, SBUX, SCG, SCHW, SE, SEE, SHW, SIAL, SJM, SLB,
SNA, SNDK, SNI, SO, SPG, SPLS, SRCL, SRE, STI, STJ, STT, STX, STZ, SWK,
SWN, SWY, SYK, SYMC, SYY, T, TAP, TDC, TE, TEG, TEL, TGT, THC, TIF,
TJX, TMK, TMO, TRIP, TROW, TRV, TSN, TSO, TSS, TWC, TWX, TXN, TXT,
TYC, UNH, UNM, UNP, UPS, URBN, USB, UTX, V, VAR, VFC, VIAB, VLO, VMC,
VNO, VRSN, VRTX, VTR, VZ, WAG, WAT, WDC, WEC, WFC, WFM, WHR, WIN,
WM, WMB, WMT, WU, WY, WYN, WYNN, XEL, XL, XLNX, XOM, XRAY, XRX,
XYL, YHOO, YUM

\subsection{Collection of Q-Q plots}
\subsubsection{Marginal Fit \label{sss:marginal_fit_qq}}

\begin{figure}
    \centering
    \includegraphics[angle=90, width=0.98\textwidth]{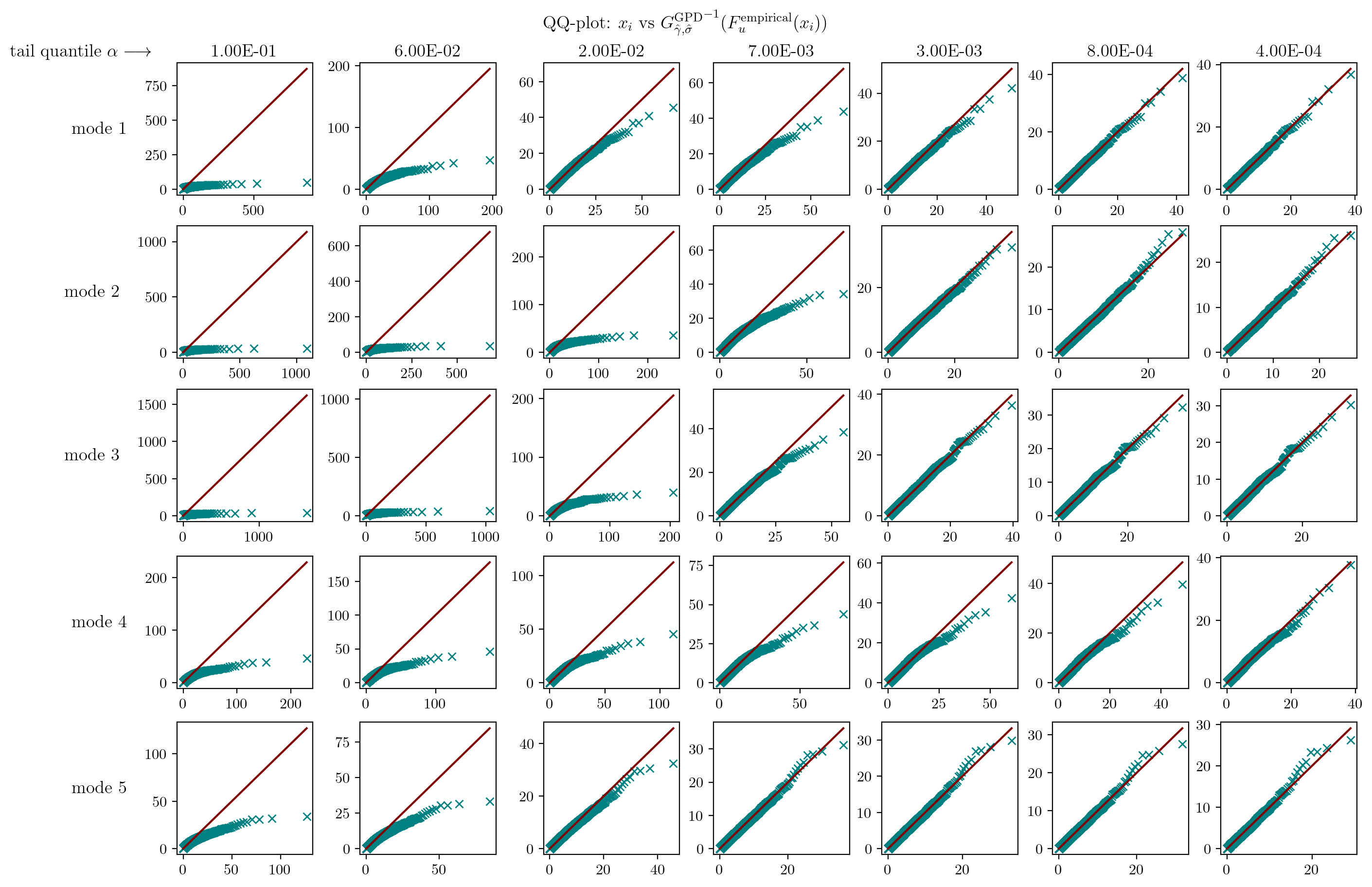}
    \caption{Q-Q plots of fitted GPD with increasing threshold, positive tail.}
    \label{fig:qq_naive+}
\end{figure}%
\begin{figure}
	\centering
	\includegraphics[angle=90, width=0.98\textwidth]{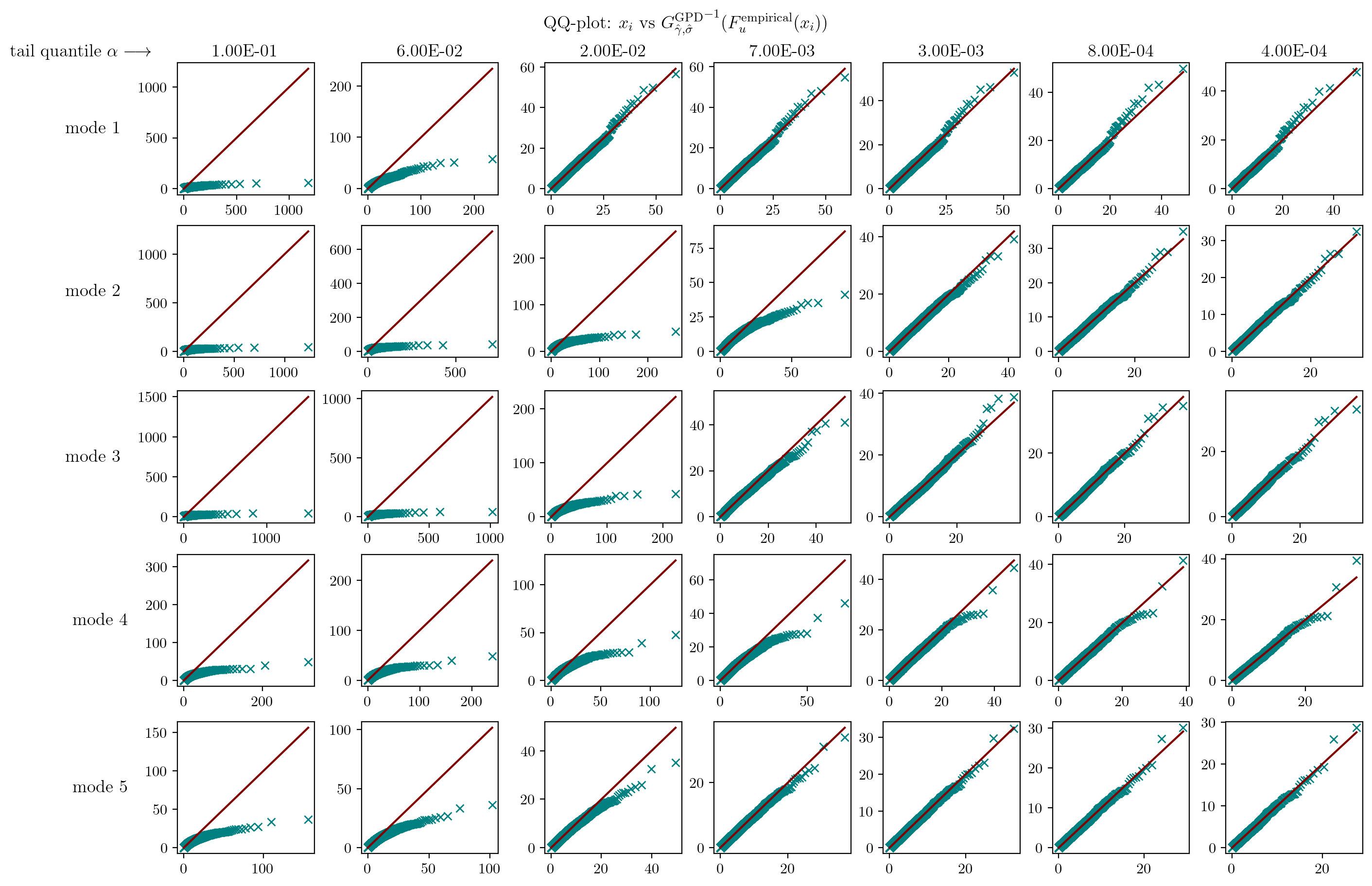}
	\caption{Q-Q plots of fitted GPD with increasing threshold, negative tail.}
	\label{fig:qq_naive-}
\end{figure}%

\FloatBarrier
\subsubsection{Residual/Local Threshold Fit\label{sss:local_fit_qq}}
\begin{figure}
    \centering
    \includegraphics[angle=90, width=0.98\textwidth]{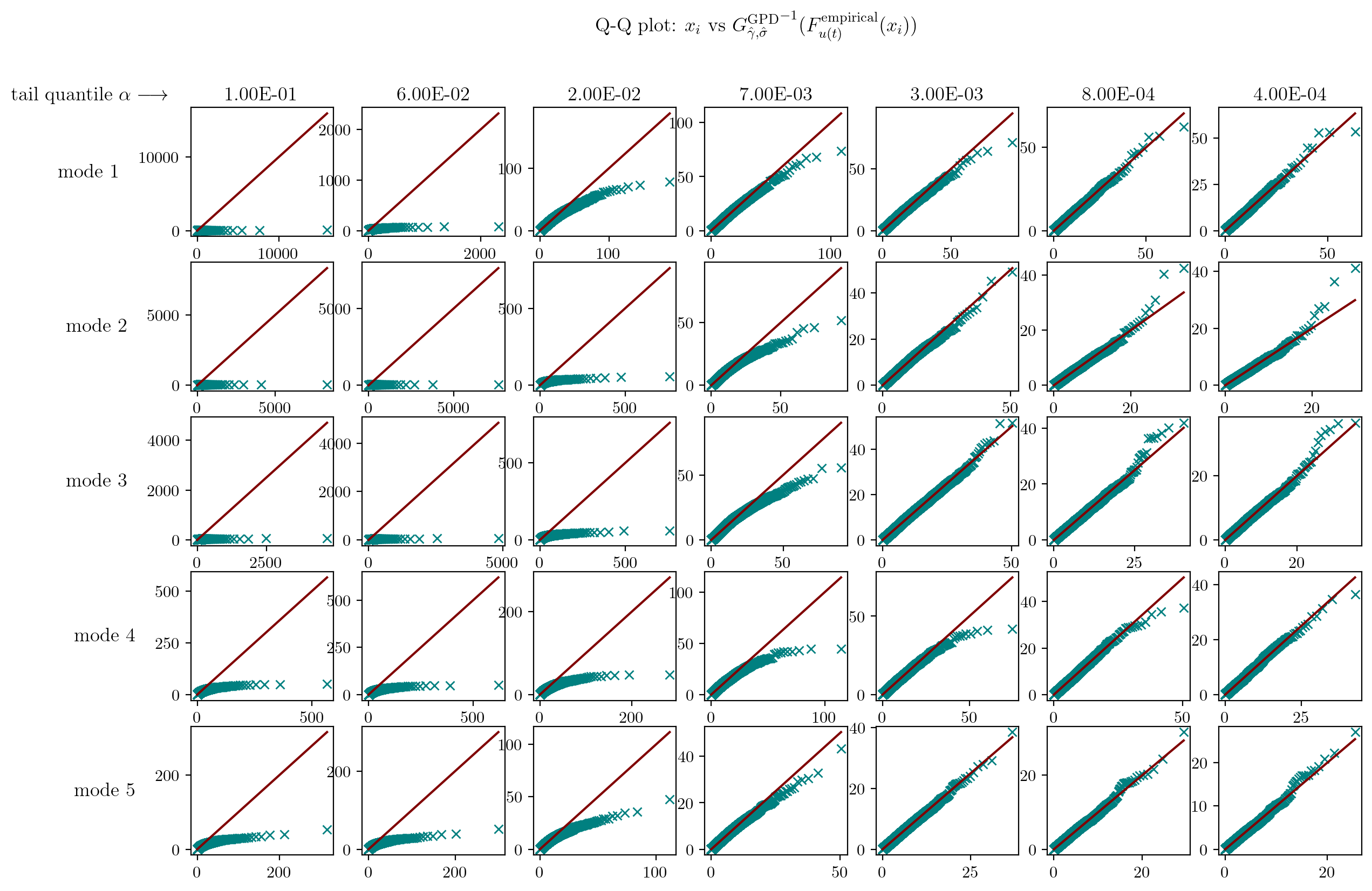}
    \caption{Q-Q plots of fitted GPD with increasing local tail-quantile threshold, positive tail.}
    \label{fig:qq_rolling+}
\end{figure}

\FloatBarrier

\subsection{Sensitivity Analysis for Local Threshold Estimation}

Regarding the analysis of Sec.~\ref{s:local_estimation}, the window length of the rolling window is an additional parameter which has to be chosen. We provide here a more in-depth analysis of the choice of this window length.\par
\begin{figure}
    \centering
    \includegraphics{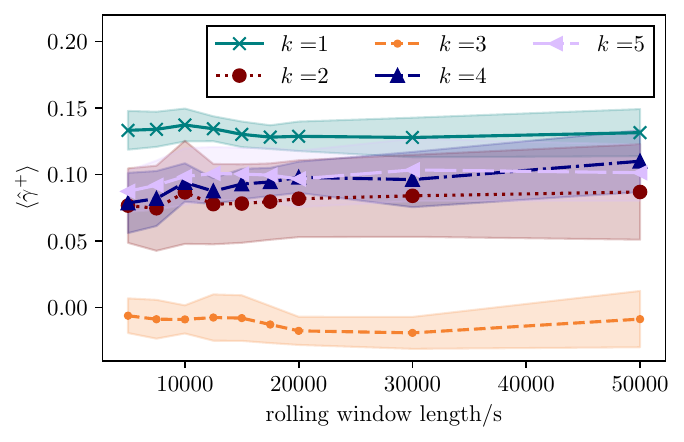}\\
    \includegraphics[]{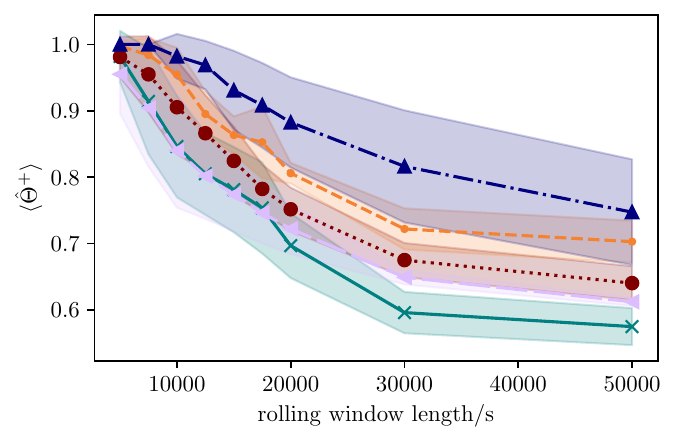}%
    \caption{Sensitivity analysis for varying rolling window length. Top/Bottom: The mean of $\hat \gamma^+$/$\hat \Theta^+$ respectively, over the ten tail-quantiles where $\rho$ was minimal, and thus the fit was the best, for varying rolling window length. Standard deviation shown as ribbons.}
    \label{fig:sensitivity_means}
\end{figure}

To keep results rather concise and manageable, we averaged over the resulting values of $\hat \gamma$ and $\hat \Theta$ in Fig.~\ref{fig:sensitivity_means}, using the ten thresholds where the fits are the best according to the NRMSD $\rho$. We show the standard deviation of these ten values as a ribbon around the calculated mean. What we find is that the mean of the tail index $\langle \gamma^+ \rangle$ is relatively stable over the window length. However, as expected, the mean of the clustering coefficient $\langle \Theta^+ \rangle$ declines with increasing window length. This indicates that extremes often occur clustered, thus the larger windows are capturing less of the nonstationarity. \par
\begin{figure}
    \centering
    \begin{subfigure}{\textwidth}
    	\centering
        \includegraphics[width=0.75\textwidth]{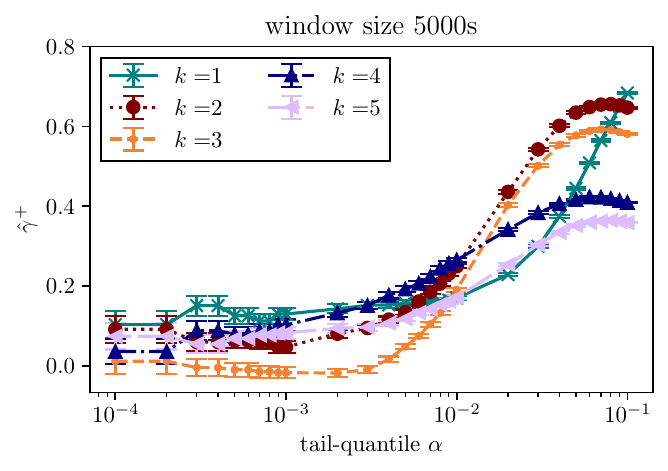}
    \end{subfigure}\\
    \begin{subfigure}{0.75\textwidth}
 \includegraphics[width=\textwidth]{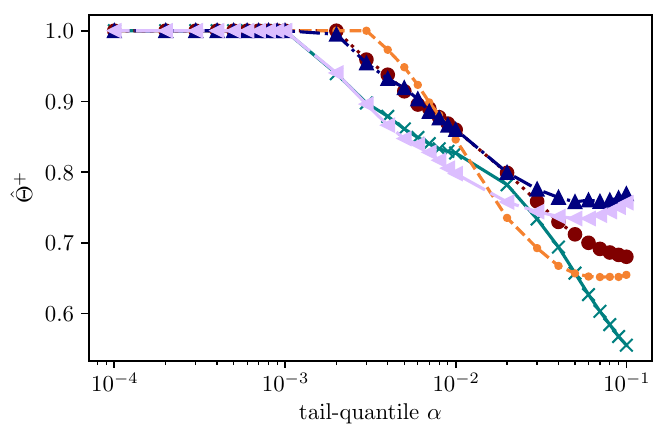}
    \end{subfigure}\\
    \begin{subfigure}{0.5\textwidth}
        \includegraphics[width=\textwidth]{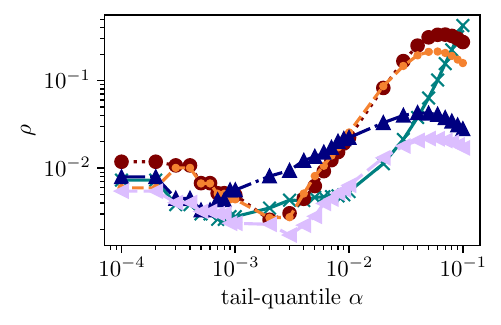}
    \end{subfigure}
    \caption{Estimation of positive tail shape parameter $\hat\gamma^+$ with a rolling threshold determined by the local tail-quantile $\alpha$ over 5000s. Top: Estimated $\hat \gamma$, standard errors according to \eqref{eq:asymptoticvariances}. Bottom left: NRMSD of estimation. Bottom right: Estimated extremal index $\hat \Theta^+$.}
    \label{fig:sensitivity_5000s}
\end{figure}

To showcase what happens at low tail-quantiles for small windows, we explicitly plot $\hat \gamma^+, \hat \Theta^+$ and $\rho$ for a window length of 5000s in Fig.~\ref{fig:sensitivity_5000s}. We see that for low and neighboring tail-quantiles, such as $1\times 10^{-4}$ and $2 \times 10^{-4}$ as well as for $3 \times 10^{-4}$ and $4\times 10^{-4}$ we get the same values for $\gamma^+$. This is due to the fact that the nearest data point to the relevant tail-quantile is not changing with the choice of the tail-quantiles anymore. This effect is unavoidable, since on average there simply are not enough data points inside the window, such that a new data point arises which exceeds the next choice of tail-quantile. \par 
In Sec.~\ref{s:local_estimation} we therefore chose to display results with a window length of 10000s, as a compromise between incorporating enough of the nonstationarity, while keeping the average number of exceedances at low tail-quantiles at a scale of one or larger.

\end{document}